\documentclass[a4paper,prl, superscriptaddress, twocolumn,longbibliography]{revtex4-2}

\usepackage{microtype}

\usepackage{silence}
\newcommand{\PRLsep}{\noindent\makebox[\linewidth]{\resizebox{0.3333\linewidth}{1pt}{$\bullet$}}\bigskip}

\usepackage[utf8]{inputenc}
\usepackage[english]{babel}
\usepackage[pdftex,dvipsnames,table]{xcolor}     
\usepackage{graphicx}
\usepackage{amsmath,amssymb}
\usepackage{amsfonts}
\usepackage{mathtools}
\usepackage{braket}
\renewcommand{\eqref}[1]{(\ref{#1})}

\usepackage{titlesec}
\titleformat*{\section}{\raggedright\bfseries\sffamily\large}
\titleformat*{\subsection}{\raggedright\bfseries\sffamily}
\titleformat*{\subsubsection}{\raggedright\bfseries\sffamily}

\usepackage{diagbox}
\usepackage{tabularray}
\usepackage{dsfont}
\UseTblrLibrary{booktabs}

\makeatletter
\renewcommand{\fnum@figure}{\textbf{Fig.~\thefigure}}
\makeatother

\usepackage{caption}
\usepackage{bm}
\renewcommand{\vec}{\ensuremath{\bm}}

\renewcommand{\Im}{\text{Im}}

\definecolor{bleu}{RGB}{0,47,108}
\usepackage{hyperref}
\hypersetup{colorlinks,citecolor={blue},linkcolor={red},urlcolor={red}}

\renewcommand{\eqref}[1]{(\ref{#1})}
\usepackage{enumitem}
\usepackage{pgfornament}
\usepackage{empheq}

\usepackage{xargs} 
\usepackage[colorinlistoftodos,prependcaption,textsize=footnotesize]{todonotes}
\newcommandx{\unsure}[2][1=]{\todo[linecolor=red,backgroundcolor=red!25,bordercolor=red,#1]{#2}}
\newcommandx{\change}[2][1=]{\todo[linecolor=blue,backgroundcolor=blue!25,bordercolor=blue,#1]{#2}}
\newcommandx{\info}[2][1=]{\todo[linecolor=OliveGreen,backgroundcolor=OliveGreen!25,bordercolor=OliveGreen,#1]{#2}}
\newcommandx{\improvement}[2][1=]{\todo[linecolor=Plum,backgroundcolor=Plum!25,bordercolor=Plum,#1]{#2}}
\newcommandx{\thiswillnotshow}[2][1=]{\todo[disable,#1]{#2}}

\usepackage{nicematrix}

\begin{document}

\title{\sffamily\Large Atomic-like selection rules in free electron scattering}
\date{\today}

\author{Simon Garrigou}
\affiliation{CEMES-CNRS, Université de Toulouse, CNRS, Toulouse, France}

\author{Hugo Lourenço-Martins}\email[]{hugo.lourenco-martins@cnrs.fr}
\affiliation{CEMES-CNRS, Université de Toulouse, CNRS, Toulouse, France}

\begin{abstract} 
Phase-shaped electron energy-loss spectroscopy (PSEELS) measures the scattering probability of structured free electron beams by a target. Over the last decade, it was shown that this scheme can be employed to emulate polarized optical spectroscopies with electrons, and therefore to transpose macroscopic optical concepts - such as dichroism - down to the deep sub-wavelength scale. In this work, we theoretically demonstrate that PSEELS can, in fact, go way further than mimicking optics and enables to map a plethora of so far inaccessible nano-optical quantities such as the electric quadrupolar momentum.
\end{abstract}

\maketitle 

Spontaneous emission refers to the process in which a quantum emitter (QE, i.e. a few-levels bound electron system) transits between two states $\ket{e}\rightarrow \ket{g}$ by emitting a photon. This free-space decay rate $\Gamma_0$ can be enhanced by weakly coupling the emitter to any type of dielectric environment  \cite{novotny_principles_2011}. Indeed, this medium increases the local density $\rho$ of electromagnetic states (EMLDOS) in which the emitter can decay into, an effect intuitively captured in the celebrated Purcell formula \cite{purcell_resonance_1946, hohenester_nano_2020}, see Fig.~\ref{fig:Figure1}(a):
\begin{equation}\label{Purcell_Factor}
    \Gamma = \frac{2 \omega}{3 \hbar \varepsilon_0} \vert\boldsymbol{d}\vert^2 \rho_{\boldsymbol{d}}(\textbf{r}_0,\omega)
\end{equation}

\noindent with $\vec{r}_0\in\mathbb{R}^3$ the position of the emitter, $\hbar$ the reduced Planck constant, $\omega$ the angular frequency of emitted radiation, $ \varepsilon_0 $ the vacuum dielectric permittivity, $\boldsymbol{d}=\braket{e \vert \hat{\vec{d}} \vert g}$ the dipole moment of the transition and $\rho_{\boldsymbol{d}}$ the EMLDOS along the dipole moment.

This whole phenomenology is not restricted to bound electron systems such as quantum dots \cite{dousse_controlled_2008}, atoms \cite{gallego_strong_2018} or molecules \cite{roslawska_mapping_2022}, but can also be observed with free electron states in a scheme called electron energy-loss spectroscopy (EELS) in a transmission electron microscope (TEM). EELS consists in analysing the weak energy-losses undergone by swift electrons ($v\sim$ 80 \% of the speed of light) during their inelastic interaction with a nano-object, thus revealing its sub-ångström dynamics, as recently demonstrated on single atoms \cite{hage_single-atom_2020}. Remarkably, a connection between the $z$-Fourier transform $\tilde{\rho}$ of the EMLDOS and the observable measured in EELS - the probability $\Gamma^{\text{EELS}}$ for the electron to lose an energy $\hbar\omega$ - was established \cite{garcia_de_abajo_probing_2008}: 
\begin{equation}\label{classical_EELS}
    \Gamma^{\text{EELS}} = \frac{2 \pi e^2}{\hbar \omega} \tilde{\rho}_z (\textbf{R}_0,q_z,\omega)
\end{equation}

\noindent where $q_z \equiv k_f-k_i=\omega/v$ is the transferred momentum, $k_i,k_f$ the initial and final electron momenta, $z$ conventionally represents the propagation axis of the electron beam, and $\vec{R}_0$ the impact parameter of the beam in its transverse plane. The similarity between equations \eqref{Purcell_Factor} and \eqref{classical_EELS} suggested a deep analogy between bound and free electron spontaneous scattering, albeit without explicitely involving any transition dipole moment - a key ingredient of the Purcell effect \eqref{Purcell_Factor}.

This difficulty was lifted over the last decade by a series of theoretical works \cite{asenjo-garcia_dichroism_2014,ugarte_controlling_2015,van_boxem_inelastic_2015} which proposed to shape \cite{verbeeck_demonstration_2018,yu_quantum_2023} and post-select \cite{grillo_measuring_2017,tavabi_experimental_2021} the electron beam wavefront. Indeed, swift electron beams being highly paraxial, this approach simply decorates the initial EELS planewave scattering problem $\ket{k_i}\rightarrow \ket{k_f}$ by the symmetry of the transverse wavefunctions $\ket{k_i}\otimes\ket{\Psi_{i,\perp}}\rightarrow \ket{k_f}\otimes\ket{\Psi_{f,\perp}}$, see Fig.~\ref{fig:Figure1}(b). This additional degree of freedom introduces new \emph{selection rules} to the transition probabilities \eqref{classical_EELS}. In particular, it was shown that the probability of transitioning from an Hermite-Laguerre-Gauss (HLG) state $\ket{\Psi}$ of order $i=1$ - representating topologically singular beams carrying one quantum of linear or orbital angular momenta \cite{bliokh_theory_2017,shen_optical_2019}  - to an unshaped Gaussian wavefront $\ket{\text{G}}$ can be written \cite{lourenco-martins_optical_2021,bourgeois_polarization-resolved_2022, bourgeois_optical_2023,nixon_inelastic_2024}:
\begin{equation}\label{pEELS}
    \Gamma^\text{pEELS} = \frac{2 \pi q_z^2}{\hbar \omega} \vert\boldsymbol{d_{\Psi}}\vert^2 \rho_{\boldsymbol{d}}(\textbf{R}_0,q_z,\omega)
\end{equation}

\noindent where $\vec{d}_\Psi=\braket{ \text{G} \vert \hat{\vec{d}} \vert \Psi}$. This scheme - referred to as polarized EELS (pEELS) -  unlocks the measurement of the EMLDOS along any polarization axis in EELS  \cite{schuler_energy-loss_2016,zanfrognini_orbital_2019,konecna_probing_2023} and completes the analogy between free electron scattering \eqref{pEELS} and the Purcell phenomenology \eqref{Purcell_Factor}.

In this letter, we demonstrate that EELS can be pushed beyond the dipole order \eqref{pEELS} by considering the scattering of HLG states of arbitrary initial $i$ and final $j$ orders with $\vert i-j\vert>1$, thus surpassing the Purcell analogy. Our formalism puts the emphasis on the quantum numbers (i.e. linear and angular momentum) conservation, therefore making selection rules explicit at \emph{all orders} and providing an intuitive picture of the scattering mechanisms. In the spirit of atomic spectroscopy, this simplicity enables us to draw look-up tables providing any experimentalist with the nano-optical quantity probed by EELS for any integer $i$ and $j$. Eventually, we close the letter with a series of three examples, first checking that both EELS and pEELS are recovered for $\vert i-j\vert=0$ and $\vert i-j\vert=1$ respectively, then demonstrating that the transition $\vert i-j\vert=2$ measures the quadrupolar component of nano-optical fields - thus showing that transitions satisfying $\vert i-j\vert>1$ access quantities unreachable with most of all-optical techniques.\\

\begin{figure}
    \centering
    \includegraphics[width=\columnwidth]{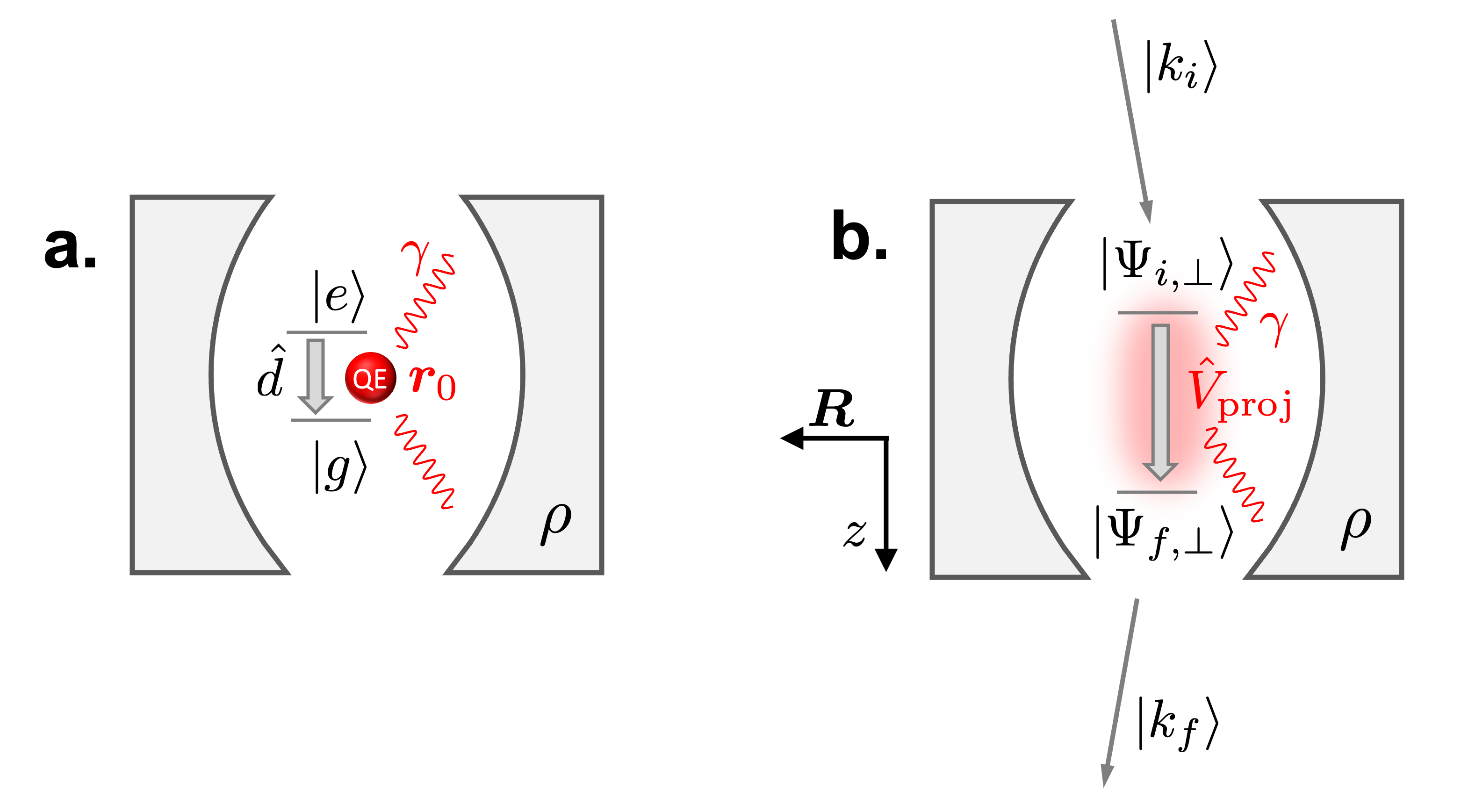}
    \caption{\textbf{Bound and free electrons spontaneous scattering.} \textbf{a.} Purcell effect: a non-trivial dielectric environment of EMLDOS $\rho$ - represented here by a cavity - enhances the spontaneous decay rate of a quantum emitter (QE). \textbf{b.} PSEELS: the exact same configuration as in a. but replacing the bound electron states by free electron orbitals i.e. planewaves $\ket{k}$ decorated by structured wavefronts $\ket{\Psi_\perp}$. $\hat{V}_\text{proj}$ the projected potential defined in \eqref{eq:proj_potential_exp}}
    \label{fig:Figure1}
\end{figure}

We begin this letter by presenting our model: the exact details of the derivation being given in the supplementary material, we will hereafter only focus on the main steps and approximations.
We start by considering a fast electron beam of longitudinal ($z$)-momentum $k_i$ and arbitrary wavefront $\Psi_{i,\perp}$. Within the paraxial approximation, the electron wavefunction can be separated between longitudinal and transverse states $\vert \psi \rangle = \vert k\rangle \otimes \vert \Psi_\perp\rangle$. 
The first step of the derivation is to compute the interaction probability between the electron beam and a target represented by a set of quantum states $\{ \ket{n}\}_{n\in\mathbb{N}}$. Here, we make no assumption on the nature of the target which can be any type of sample (e.g. phononic, photonic or plasmonic material, molecule, van der Waals material, etc), we will always describe it as a polariton field \cite{basov_polaritons_2016,rivera_lightmatter_2020,kfir_optical_2021}, and thus generically denotes them as \emph{photonic states}. We moreover impose a set of common approximations valid for swift electron scattering \cite{garcia_de_abajo_optical_2010}: Quasi-static ($c \rightarrow +\infty$), single scattering (weak interaction), non-recoil (large electron velocity $v$) and local potential (narrow beam waist) approximations. Within this framework, one can show that the electron-target interaction potential reduces to a projected potential: 
\begin{equation}\label{eq:proj_potential_exp}
    \hat{V}_\text{proj} (q_z,\omega) = \int dz \langle z \vert \hat{V} \vert z \rangle e^{i q_z z}
\end{equation}

\noindent where the longitudinal degree of freedom of the electron has been absorbed in the definition of an effective projected potential $\hat{V}_\text{proj}$ \cite{findlay_modeling_2008,dwyer_multislice_2005,coene_inelastic_1990,lubk_jacobs_2015}, translating the delocalization of interaction along the electron path, see Fig. \ref{fig:Figure1}(b). The interaction probability is then given by a direct application of the Fermi's golden rule:
 \begin{equation} \label{eq:EELS_qproba_g0}
 \begin{split}
    \Gamma^\text{PSEELS}(\omega) & =   \tilde{\sum_n}  \left| \frac{1}{\hbar v}  \langle \Psi_{f,\perp} \vert \langle n \vert \hat{V}_\text{proj} \vert 0 \rangle \vert \Psi_{i,\perp} \rangle \right|^2\\
    &= \tilde{\sum_n} | g_{0,n} |^2
     \end{split}
\end{equation}

\noindent where the energy conservation is encompassed in the new definition of the sum $\tilde{\sum}_n = \sum_n \delta(\omega_0 - \omega_n - \omega) $. One can see that, when preparing $\vert\Psi_{i,\perp}\rangle$ and post-selecting $\vert\Psi_{f,\perp}\rangle$ transverse electron states (see Fig.~\ref{fig:Figure1}(d)) - a scheme denoted as phase-shaped EELS (PSEELS) - one selects a specific transition occurring between discrete electronic states and for all accessible photonic states $\vert n \rangle$, represented by the coupling constant $g_{0,n}$. The total loss-probability \eqref{eq:EELS_qproba_g0} is then given by the sum over all the possible final photonic states. For pedagogy, in the rest of the paper, we will drop this sum and focus on one specific transition in the target $g_{0,n}$. From \eqref{eq:EELS_qproba_g0}, the discrete electron transverse states essentially behave as atomic orbitals, imposing specific selection rules by their symmetry. However, while atomic orbitals are three dimensional objects with shapes fixed by the nature of the atom, \emph{free electron orbitals} (i.e $\vert\Psi_{i,\perp}\rangle$ and $\vert\Psi_{f,\perp}\rangle$) are two dimensional and their shape can be chosen arbitrarily through beam shaping and post-selection.\\

In order to explicit the connection between free electron orbitals and atomic orbitals, as well as the corresponding selection rules, we will now perform a multipolar development of the projected potential. Indeed, since the electron beam waist (Angstrom) is small compared to the electrostatic potential variations scale (nanometer), one can consider the projected potential and its derivative as constant over the electron wavefunction, and represented by the function $\hat{V}_\text{proj}(\textbf{R}_0,q_z,\omega)$, where $\vec{R}_0$ is the impact parameter of the electron in the transverse plane. In this limit, one can perform a multipolar developement equivalent to the Power-Zienau-Woolley form of the minimal-coupling \cite{novotny_principles_2011}:
\begin{equation}\label{eq:projected_potential_mult_dev}
    \hat{V}_\text{proj} (q_z, \omega) = \underbrace{\hat{\rho} \hat{\Phi} (q_z, \omega)}_{\hat{V}_0} - \underbrace{\hat{\textbf{d}} \cdot \hat{\textbf{E}} (q_z, \omega)}_{\hat{V}_1} - \underbrace{\hat{\overleftrightarrow{Q}} \nabla \cdot \hat{\textbf{E}} (q_z, \omega)}_{\hat{V}_2} + ...
\end{equation}

\noindent where the term involving the k$^\text{th}$ derivative of the potential will be noted $\hat{V}_k$. As we will confirm further, the term $\hat{d} $ in $\hat{V}_1$ is the dipolar operator at the origin of \eqref{pEELS}. Now, each term of $\hat{V}_\text{proj}$ is composed of two operators, acting either on the electron transverse degree of freedom or the photonic degrees of freedom. Plugged back into \eqref{eq:EELS_qproba_g0}, the latter gives expectation values of the type $\langle n \vert \hat{A} \vert 0 \rangle =A_{n0} $, with $\hat{A}$ a field operator, e.g $\hat{\textbf{E}}$ or $ \hat{\Phi}$. In later developments, we will focus on a single $n$ photonic mode of the system and omit the $n0$ subscript for clarity.

The last step of our derivation consists in giving an alternative and more intuitive form to the electron i$^\text{th}$ momentum operators $\hat{\rho}, \hat{\textbf{d}}, \hat{\overleftrightarrow{Q}}, \hdots$. This is done by describing the transverse electron field with a ladder operator formalism typically encountered in the two dimensional quantum harmonic oscillator. Introducing the annihilation $\hat{a}_u$ and creation $\hat{a}_u^\dag$ operators along direction $u$. With $u \in \{x,y\} \text{ or } u\in \{\circlearrowleft,\circlearrowright\}$, linear and circular basis respectively. The electronic position operators can be re-written as \cite{claude_cohen-tannoudji_cohen-tannoudji_2019}:
    \begin{empheq}[left={\empheqlbrace\,}]{align}
    \hat{x} = \frac{w_0}{2} \left( \hat{a}_x^{} + \hat{a}_x^\dag \right) = \frac{w_0}{2\sqrt{2}} \left( \hat{a}_\circlearrowright^{} + \hat{a}_\circlearrowleft^{} + \hat{a}_\circlearrowright^\dag + \hat{a}_\circlearrowleft^\dag \right)\\
    \hat{y} = \frac{w_0}{2} \left( \hat{a}_y^{} + \hat{a}_y^\dag \right) = \frac{i w_0}{2\sqrt{2}} \left( \hat{a}_\circlearrowleft^{} - \hat{a}_\circlearrowright^{} - \hat{a}_\circlearrowleft^\dag + \hat{a}_\circlearrowright^\dag \right)
    \end{empheq}

With the circular operators $\hat{a}_\circlearrowleft^{}= \frac{1}{\sqrt{2}} (\hat{a}_x^{} - i \hat{a}_y^{} )$ and $\hat{a}_\circlearrowright^{}= \frac{1}{\sqrt{2}} (\hat{a}_x^{} + i \hat{a}_y^{} )$. Hence, the dipolar term of the interaction potential in linear basis now simply reads:
\begin{equation}\label{Dipolar_moment_Hermite}
    -\hat{\textbf{d}}.\vec{E} = \frac{e w_0}{2} \left[ (\hat{a}_x^{} + \hat{a}_x^\dag) E_{x} +  (\hat{a}_y^{} + \hat{a}_y^\dag) E_{y} \right]
\end{equation}

\noindent Note that here, since the photonic degrees of freedom have been absorbed in $\vec{E}$, the previous expression corresponds to a pure transverse electron operator. Every term (quadrupolar, octupolar...) of \eqref{eq:projected_potential_mult_dev} is now expressed as a sum/product of these ladder operators. Remarkably, one can show that each term $\hat{V}_k$ of the interaction potential \eqref{eq:projected_potential_mult_dev} only involves products $\hat{a}^{(k)}$ of $k$ ladder operators, annihilation and/or creation along directions $u \in \{x,y,\circlearrowleft,\circlearrowright\}$:
\begin{equation}
     \braket{n\vert\hat{V}_\text{proj}\vert 0}=\sum_{k} \braket{n\vert\hat{V}_k[\hat{a}^{(k)}]\vert 0}
\end{equation}

\noindent The integer $k$ will be called the order of the transition in the following. For instance, the term \eqref{Dipolar_moment_Hermite} represents a transition of order 1 where ladder operators only appear to the power one.

Now, why is this new form of the electron operators in terms of ladder operator advantageous? As mentioned in the introduction, the vast majority of electron beam wavefronts considered in both the experimental and theoretical literature are Hermite- (HG) and Laguerre-Gauss (LG) states. The key property of these sets of states is that they also form two eigenbasis of the 2D quantum harmonic oscillator. Thus, as demonstrated in the pioneering work of G. Nienhuis and L. Allen \cite{nienhuis_paraxial_1993}, all the properties of the HG and LG light beams can be expressed in term of ladder operators, in complete analogy with the quantum harmonic oscillator. Here, we reuse this technique in our model to describe not light but electron HLG beams, the method being essentially the same. In particular, every electron transverse state can be built by the successive application of ladder operators on a simple Gaussian state $\vert G \rangle$: 
    \begin{empheq}[left={\empheqlbrace\,}]{align}
      \vert HG_{n_x,n_y}^{} \rangle &= \frac{1}{\sqrt{n_x! n_y!}} (\hat{a}^\dag_x)^{n_x} (\hat{a}^\dag_y)^{n_y} \vert G \rangle \label{eq:HG_ladder_op}\\
      \vert LG_{n_\circlearrowright,n_\circlearrowleft}\rangle &= \frac{1}{\sqrt{n_\circlearrowright! n_\circlearrowleft!}} (\hat{a}_\circlearrowright^\dag)^{n_\circlearrowright}(\hat{a}_\circlearrowleft^\dag)^{n_\circlearrowleft} \vert G \rangle \label{eq:LG_ladder_op}
    \end{empheq}

In this text, we use the notation $\vert LG_{n_\circlearrowleft,n_\circlearrowright} \rangle$ which deviates from common $\vert LG_p^l \rangle$, the correspondence being given by $p = \min(n_\circlearrowleft,n_\circlearrowright)$ and $l = n_\circlearrowleft -n_\circlearrowright$. In the exact same way as we did for the interaction potential, we define the order $i$ of a pure HG or LG beam profile as the number of creation operators required to build it from the Gaussian state. Any transverse state can be transformed into another transverse state by application of ladder operators. Then, the order $|i-j|$ of the transition between two transverse states $i$ and $j$ is defined as the number of operators needed to transit from $i$ to $j$. A first order transition corresponds to $|HG_{1,0}\rangle \rightarrow |G\rangle$ or $|LG_{1,0}\rangle \rightarrow |LG_{1,1}\rangle$ for instance. 

At this stage, the strength of this approach starts to be visible: both the multipolar interaction potential and the transverse electron states are now expressed through the \emph{same} ladder operators: computing transition probability now only relies on simple Hilbert space algebra. But the deep interest of this model is revealed when calculating any transition between two e.g. HG states of order $i$ and $j$, one gets to leading order:
\begin{equation} \label{Amplitude_HG_general}
    g_{0,i\rightarrow j} = \sum_{k} \mathcal{M}^{(k)}[\hat{V}_k]\;\delta_{k,\vert i-j \vert}
\end{equation}

\noindent where $\mathcal{M}^{(k)}$ corresponds to the transition of order $i$ involving only the multipolar term $\hat{V}_k$. The analytical expressions of the amplitudes $\mathcal{M}^{(k)}$ can be quite complex - especially for increasing order $k$ - and are exactly derived in the supplementary material. However, one can already notice that, remarkably, the interaction probability will be non-zero if and only if the order of the transition in the transverse electron state matches the order of a multipolar term of the projected potential, expressed through the condition $\delta_{k,\vert i-j \vert}$, with $\delta_{\mu,\nu}$ the Kronecker delta. 
The latter is reminiscent of the emergence of selection rules and is responsible for a pairing between the order of the transition $|i-j|$ and a specific multipolar component of the field, as shown on tables \ref{tab:table1} and \ref{tab:table2}. We visualize here that different experiments sharing the same $|i-j|$ probe in fact the same quantity, the deciding factor being the nature of the transition more than that of the initial and final state.

This pairing is not accidental but instead a consequence of the Noether theorem. For instance, consider a transition between two HG states with $\Delta n_x=1$ and $\Delta n_y = 0$. Then, the corresponding amplitude will scale as $g_{0, n_xn_y \rightarrow n_x'n_y'}^{HG} \propto E_{x}$. Keeping in mind that $n_x$ represents a quantum of linear momentum in the direction $x$, the selection rule expresses the fact that if the electron interacts with a field oriented along the $x$-direction, it must undergo a transverse momentum kick in the same direction.\\

\begin{table}[ht]
    \caption{$\vert \Psi_{f,\perp}\rangle \rightarrow \vert \Psi_{i,\perp} \rangle$ PSEELS transition amplitudes for Hermite-Gauss beam shapes. The amplitudes are given only as the probed nanophotonic quantity, for brevity we remove their dependency $(\textbf{R}_0,q_z,\omega)$. Beam profiles are plotted in amplitude of the wavefunction, red for positive, blue for negative. $(n_x,n_y)$ index of the wavefront is displayed in the bottom left corner.}
    \begin{tblr}{ 
      colspec = {X[c,h]X[c]X[c]X[c]},
      stretch = 0,
      rowsep = 0pt,
      hlines,
      vlines,
    }

     \diagbox[width=5.5em, height=3\line]{$\vert \Psi_{f,\perp}\rangle$}{$\vert \Psi_{i,\perp} \rangle$} & 
      \includegraphics[height=1cm]{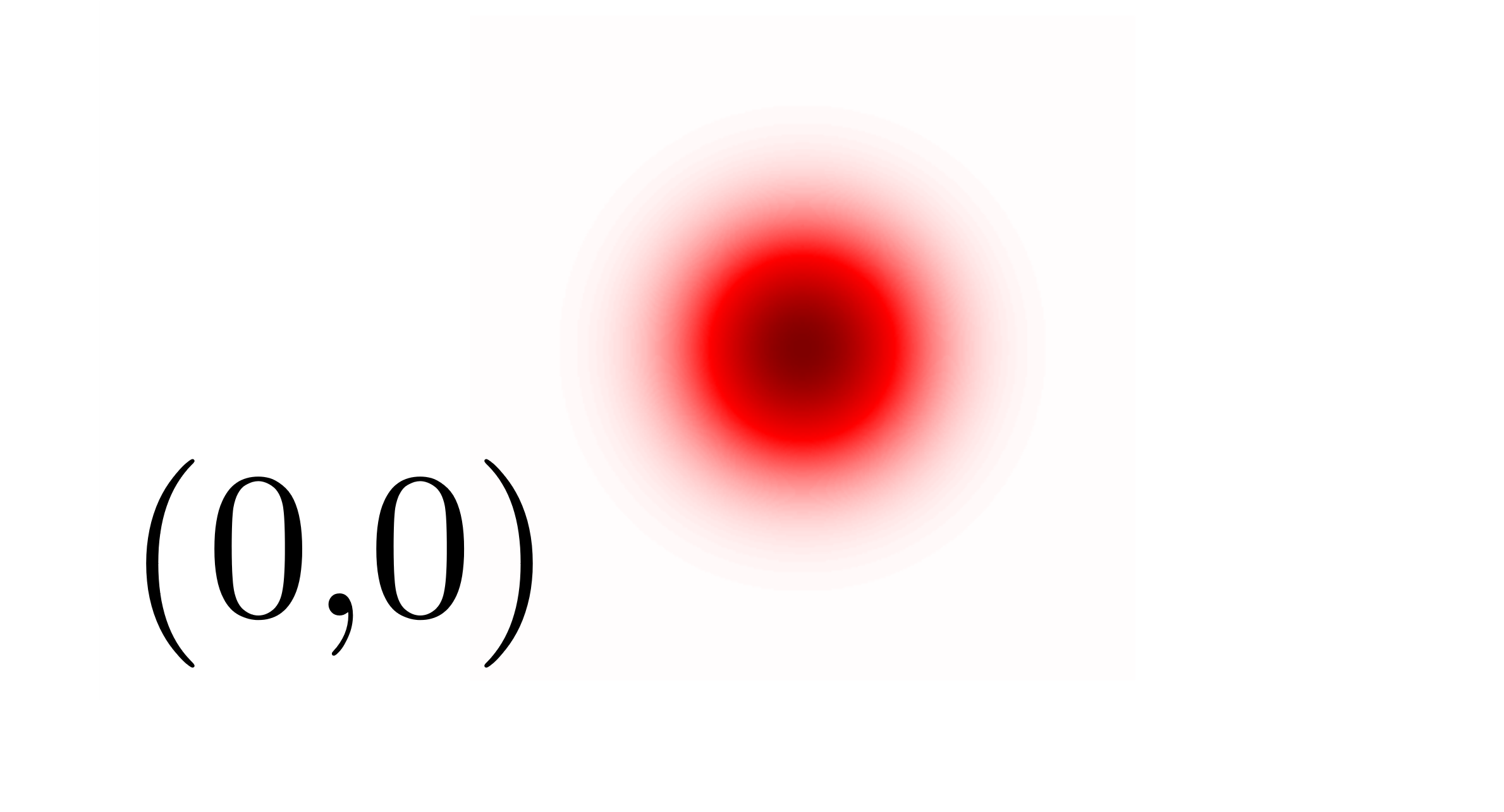} & \includegraphics[height=1cm]{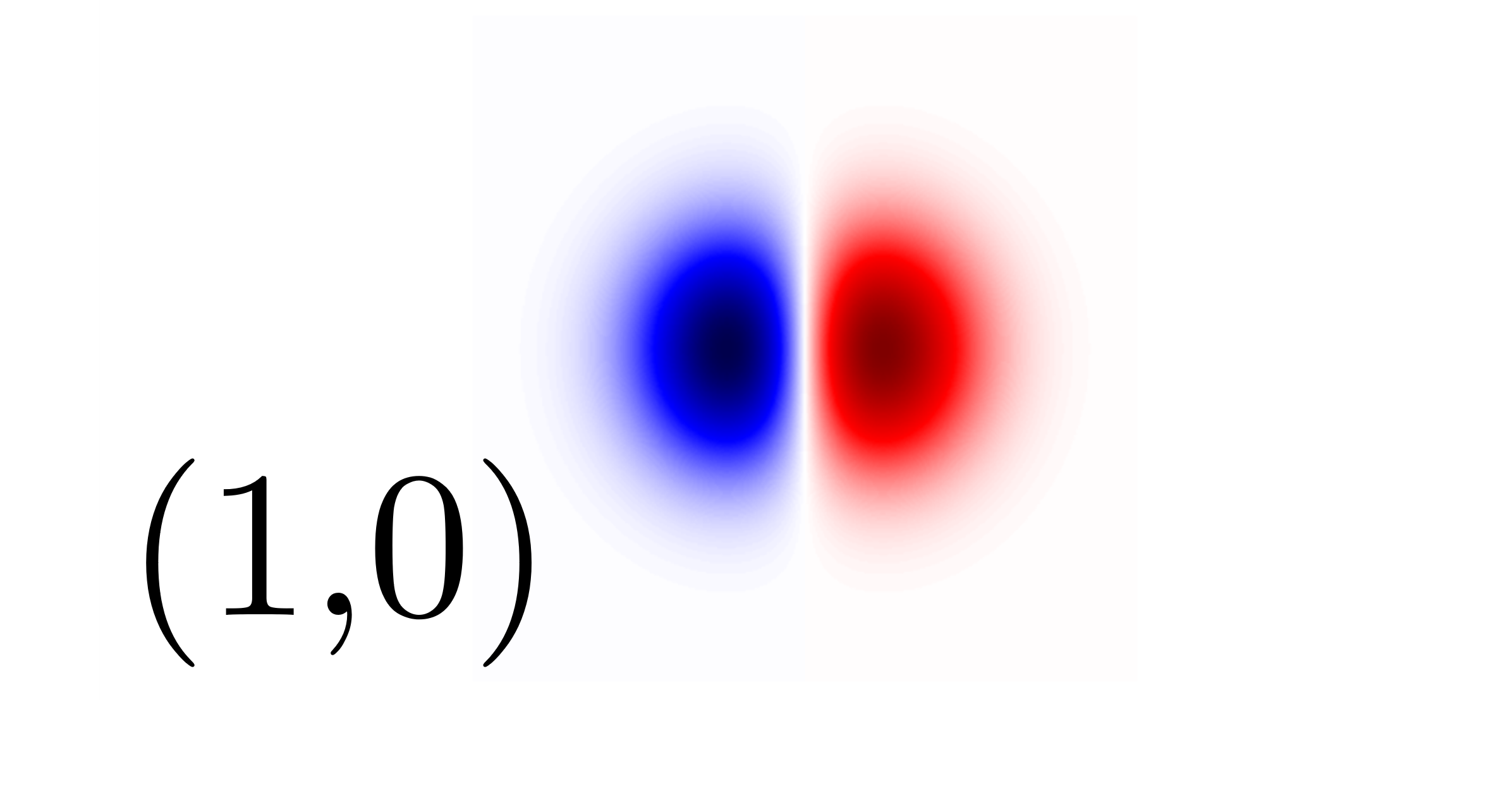}  & \includegraphics[height=1cm]{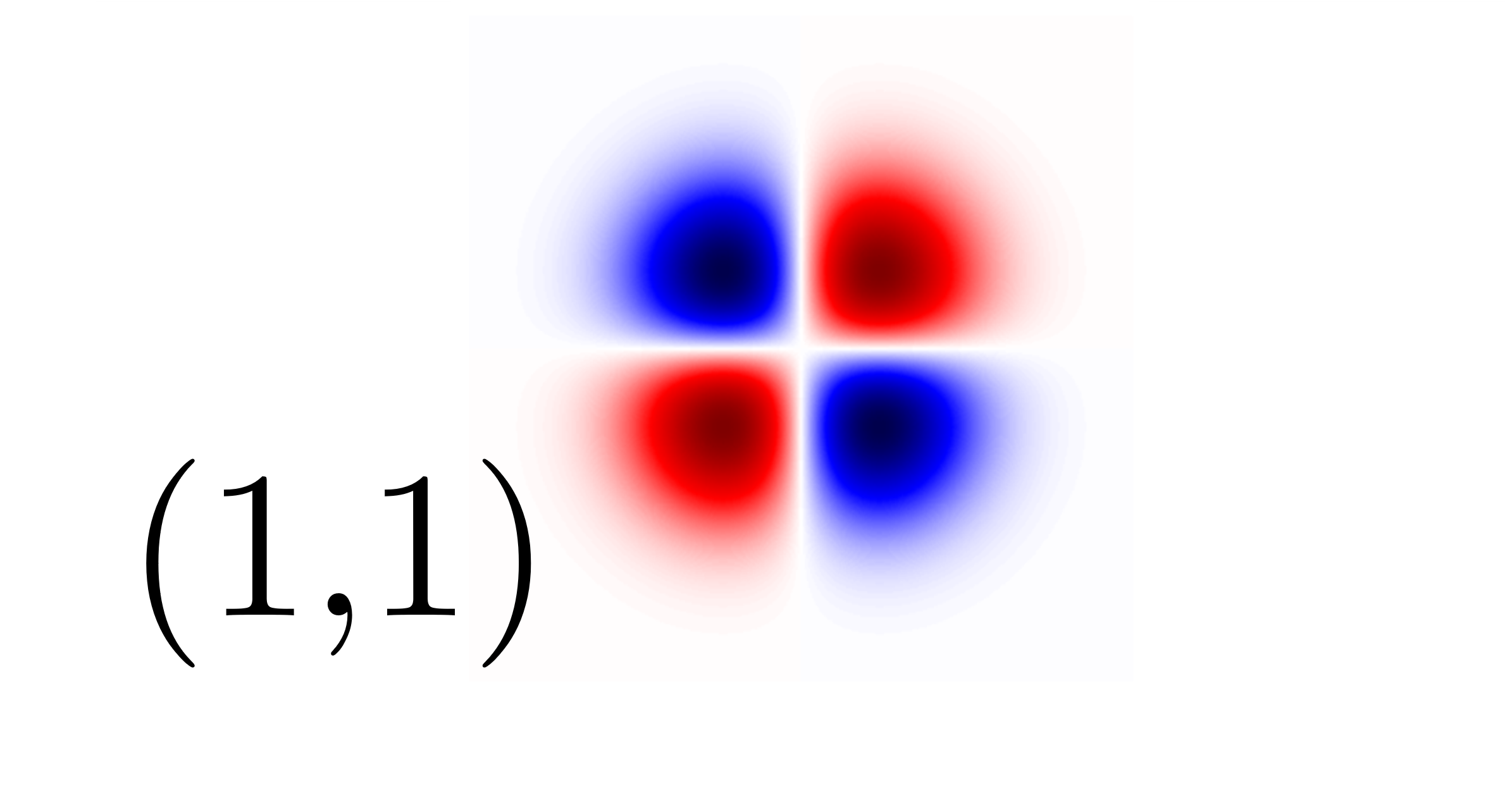} \\
     \includegraphics[height=1cm]{G.png} & \SetCell{bg=yellow!40} $\Phi_{}$ &\SetCell{bg=blue!20} $E_{x}$ &\SetCell{bg=green!10} $\partial_x E_{y} + \partial_y E_{x}$ \\ 
     \includegraphics[height=1cm]{HG10.png} &\SetCell{bg=blue!20}  $E_{x}$ & \SetCell{bg=yellow!40} $\Phi_{}$ &\SetCell{bg=blue!20} $E_{y}$ \\ 
     \includegraphics[height=1cm]{HG11.png} &\SetCell{bg=green!10} $\partial_x E_{y} + \partial_y E_{x}$  &\SetCell{bg=blue!20} $E_{y}$ & \SetCell{bg=yellow!40} $\Phi_{}$ \\
     \includegraphics[height=1cm]{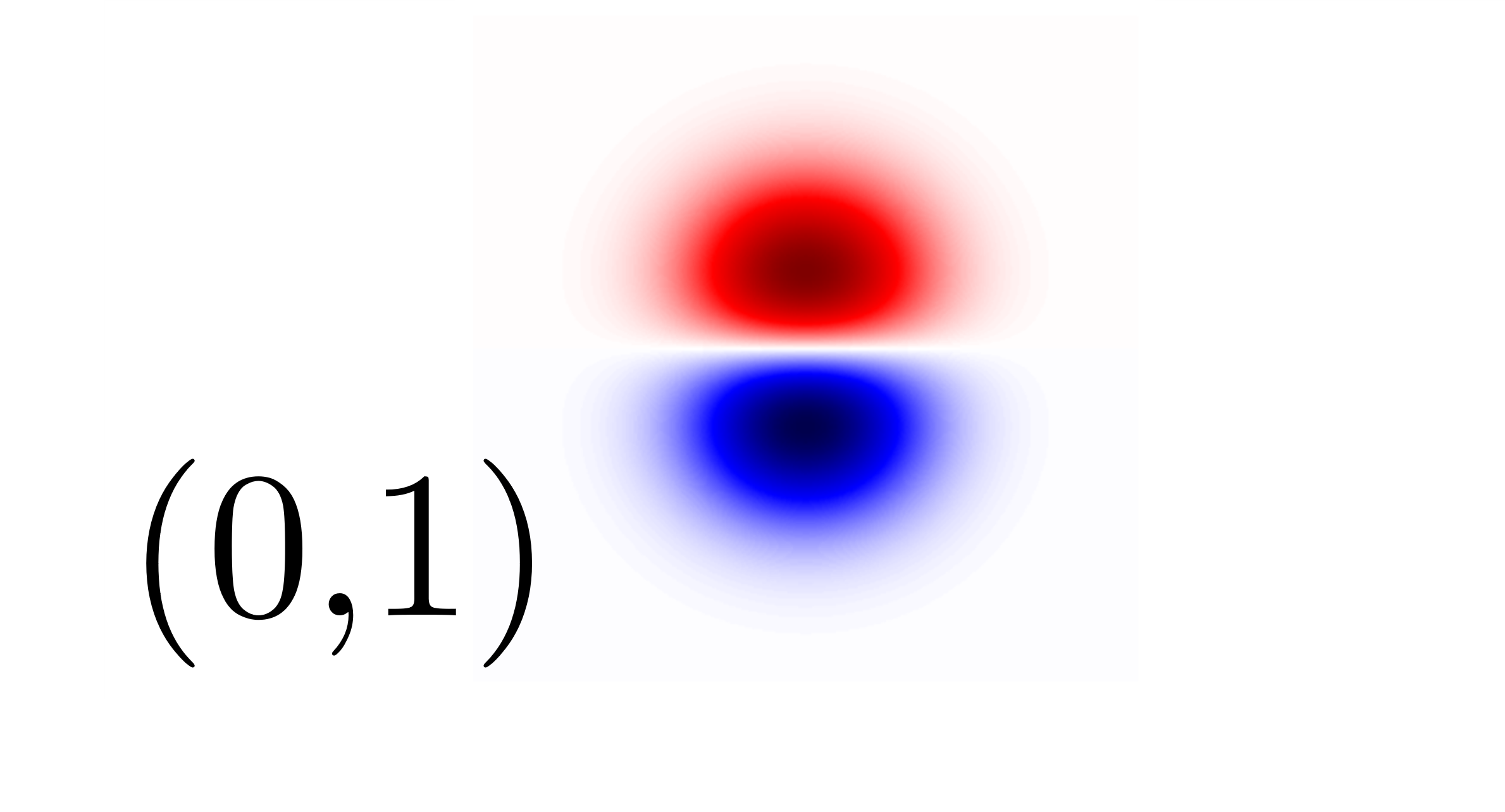} &\SetCell{bg=blue!20} $E_{y}$  &\SetCell{bg=green!10} $\partial_x E_{y} + \partial_y E_{x}$ &\SetCell{bg=blue!20} $E_{x}$ \\ 

     \includegraphics[height=1cm]{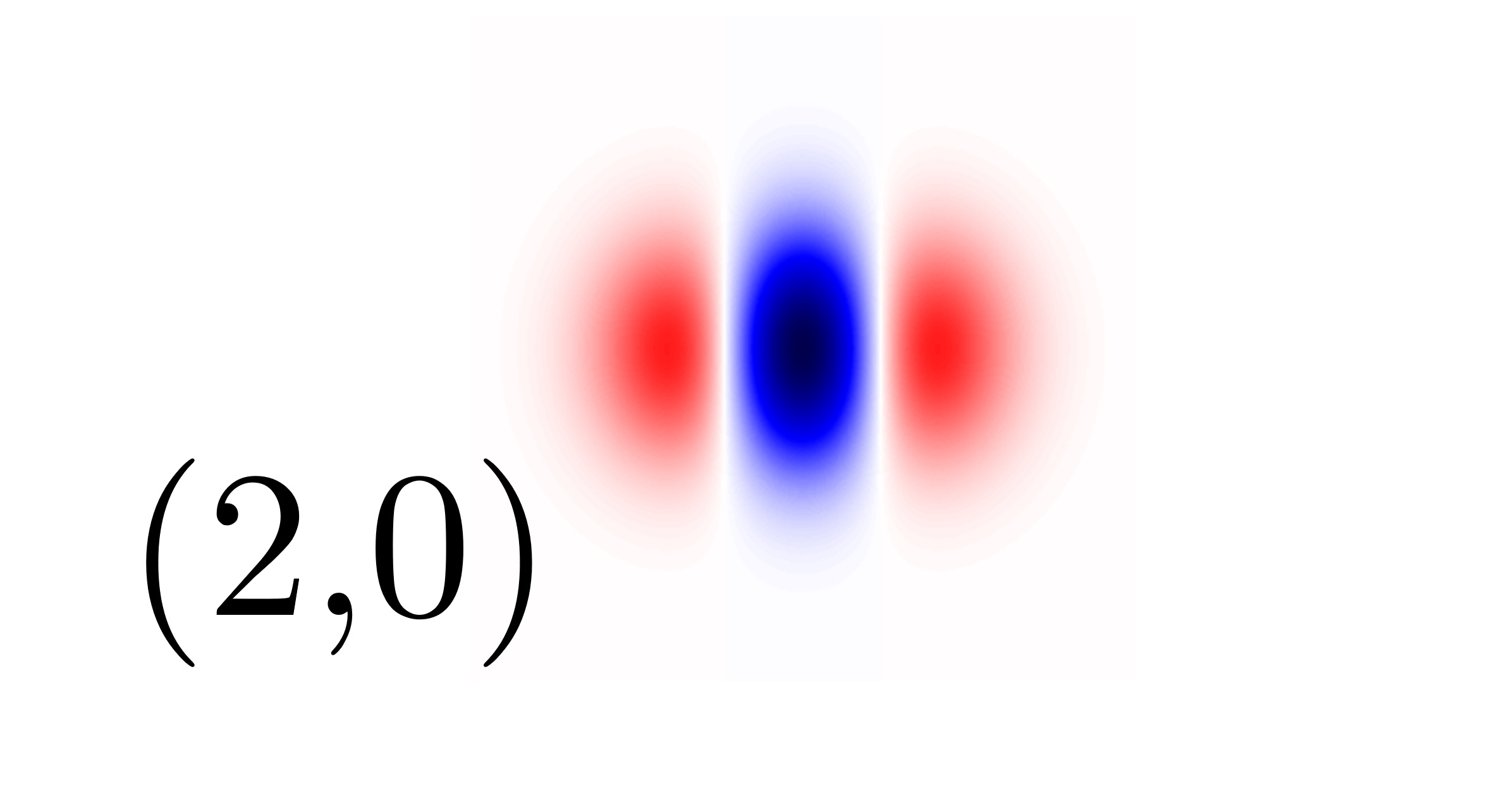} &\SetCell{bg=green!10} $\partial_x E_x$ &\SetCell{bg=blue!20}  $E_{x}$ & \SetCell{bg=green!10}$\partial_x E_{y} + \partial_y E_{x}$ \\
    \end{tblr}
    \label{tab:table1}
\end{table}    

\begin{table}[ht]
    \caption{$\vert \Psi_{f,\perp}\rangle \rightarrow \vert \Psi_{i,\perp} \rangle$ PSEELS transition amplitudes for Laguerre-Gauss beam shapes. The amplitudes are simplified to the probed nanophotonic quantity, for brevity we remove their dependency $(\textbf{R}_0,q_z,\omega)$. Beam profiles are plotted in the squared norm of the wavefunction, the orbital angular momentum of the mode is displayed in the upper left corner and $(n_\circlearrowleft,n_\circlearrowright)$ index of the wavefront is displayed in the bottom left corner.}
    \begin{tblr}{ 
      colspec = {X[c,h]X[c]X[c]X[c]},
      stretch = 0,
      rowsep = 0pt,
      hlines, 
      vlines,
    }

     \diagbox[width=5.5em, height=3\line]{$\vert \Psi_{f,\perp}\rangle$}{$\vert \Psi_{i,\perp} \rangle$} & \includegraphics[height=1cm]{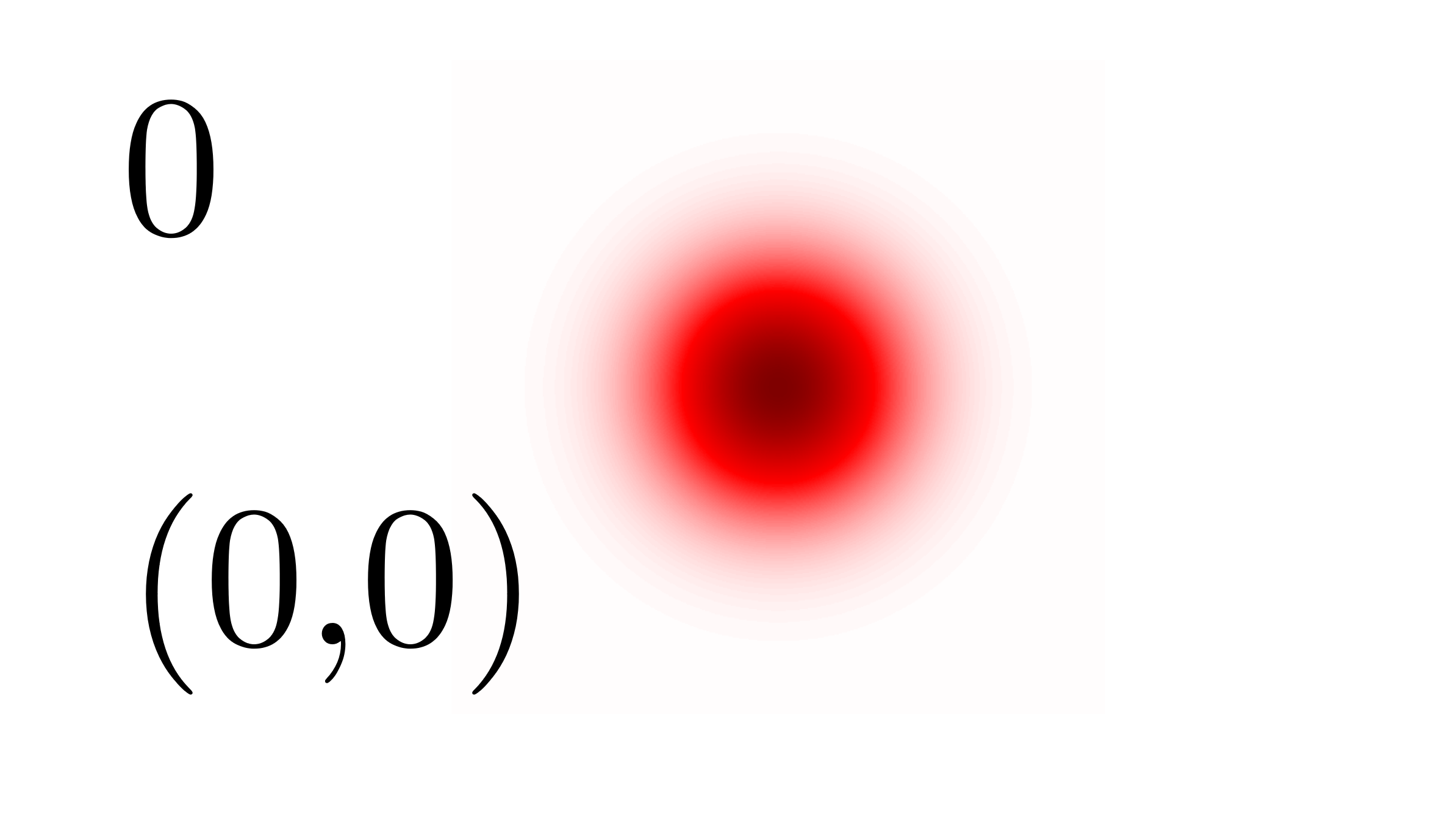} & \includegraphics[height=1cm]{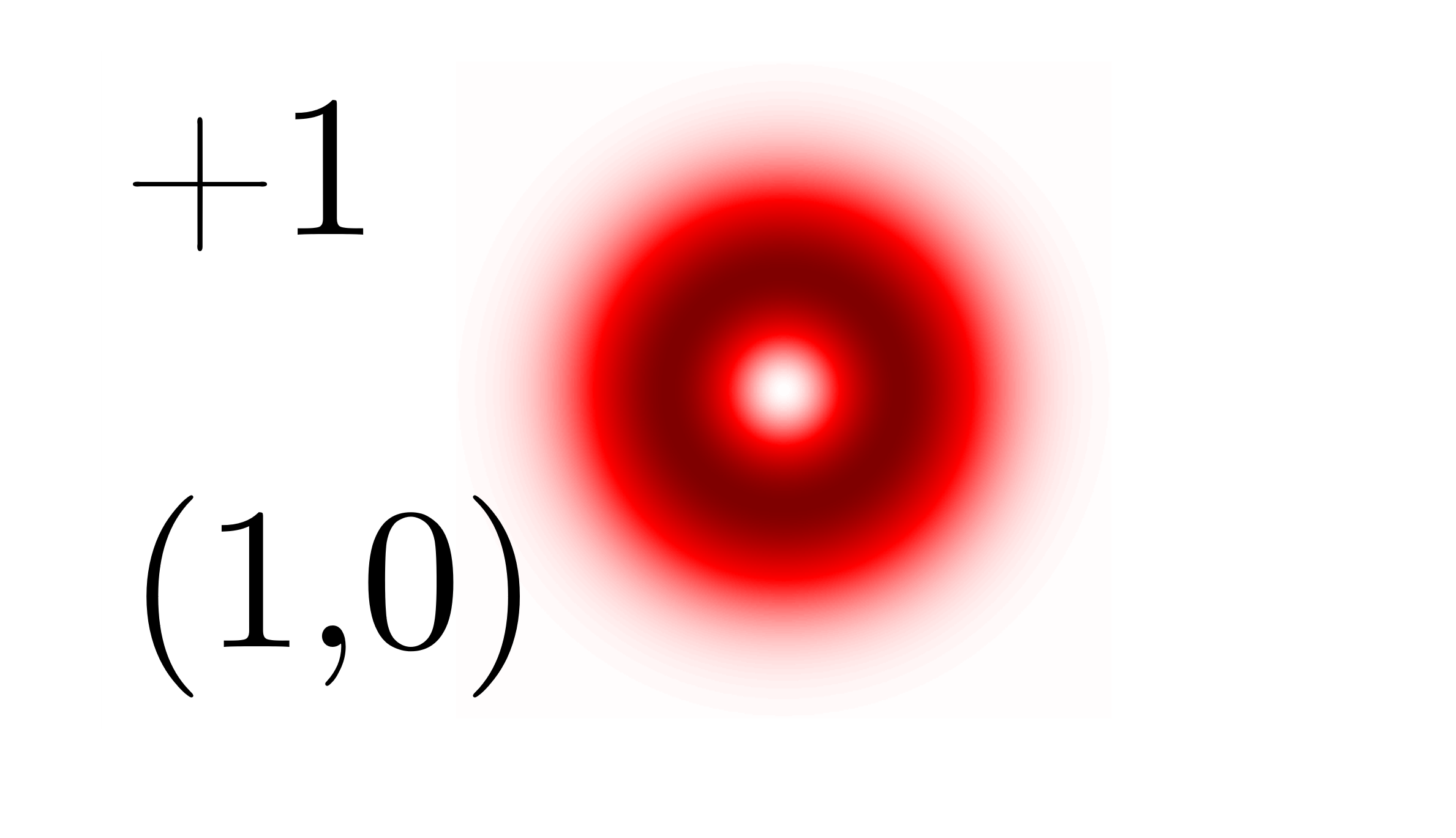} & \includegraphics[height=1cm]{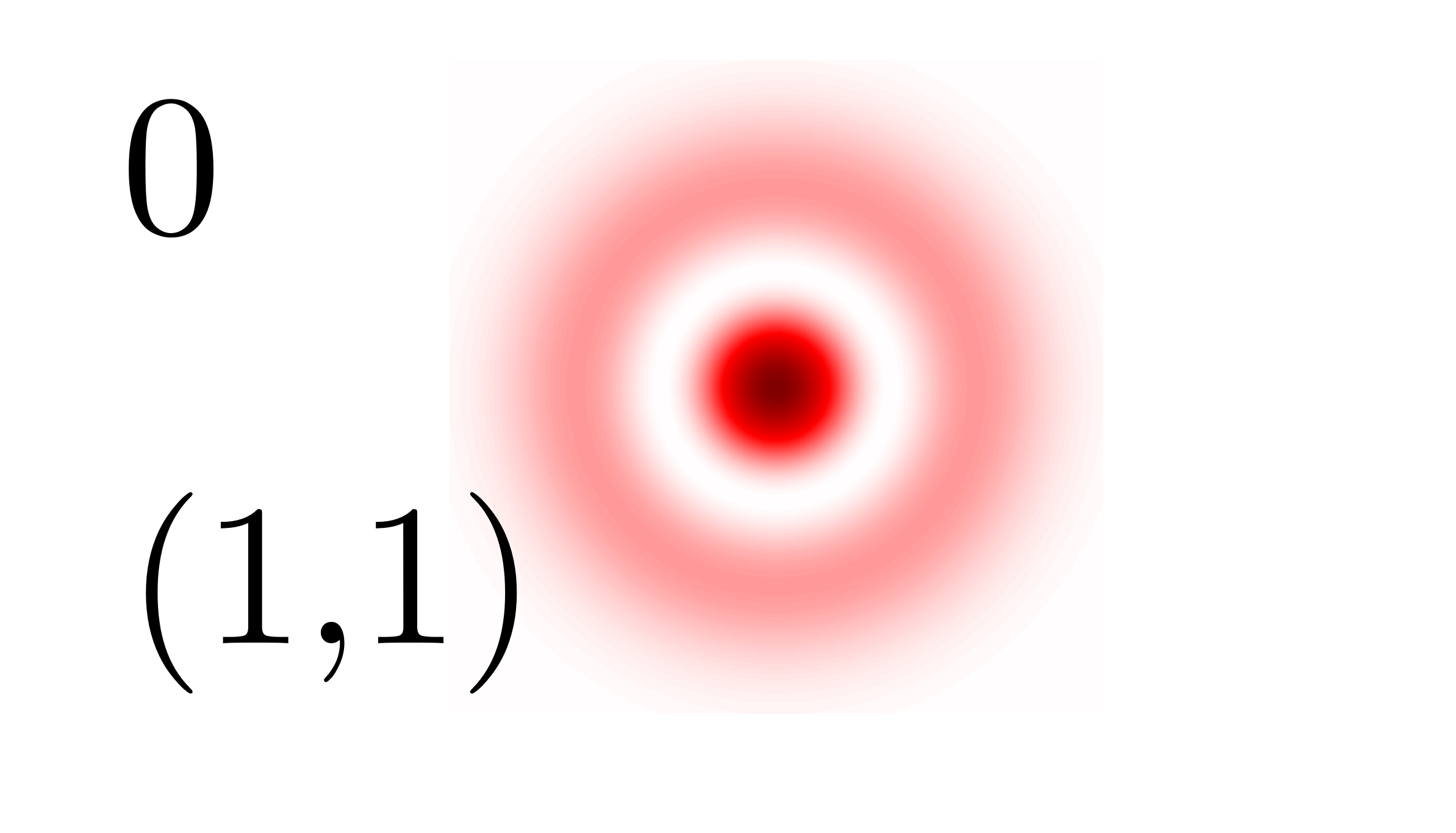} \\
     \includegraphics[height=1cm]{LG_0_0.png} &\SetCell{bg=yellow!40} $\Phi_{}$ &\SetCell{bg=blue!20} $E_{\circlearrowleft}$ &\SetCell{bg=green!10} $\partial_\circlearrowright E_{\circlearrowright} +\partial_\circlearrowleft E_{\circlearrowleft}$ \\ 
     \includegraphics[height=1cm]{LG_0_1.png} &\SetCell{bg=blue!20} $E_{\circlearrowright}$ & \SetCell{bg=yellow!40}$\Phi_{}$ &\SetCell{bg=blue!20} $E_{\circlearrowleft}$ \\ 
     \includegraphics[height=1cm]{LG_1_0.png} &\SetCell{bg=green!10} $\partial_\circlearrowright E_{\circlearrowright} +\partial_\circlearrowleft E_{\circlearrowleft}$ &\SetCell{bg=blue!20} $E_{\circlearrowleft}$ &\SetCell{bg=yellow!40} $\Phi_{}$ \\ 
     \includegraphics[height=1cm]{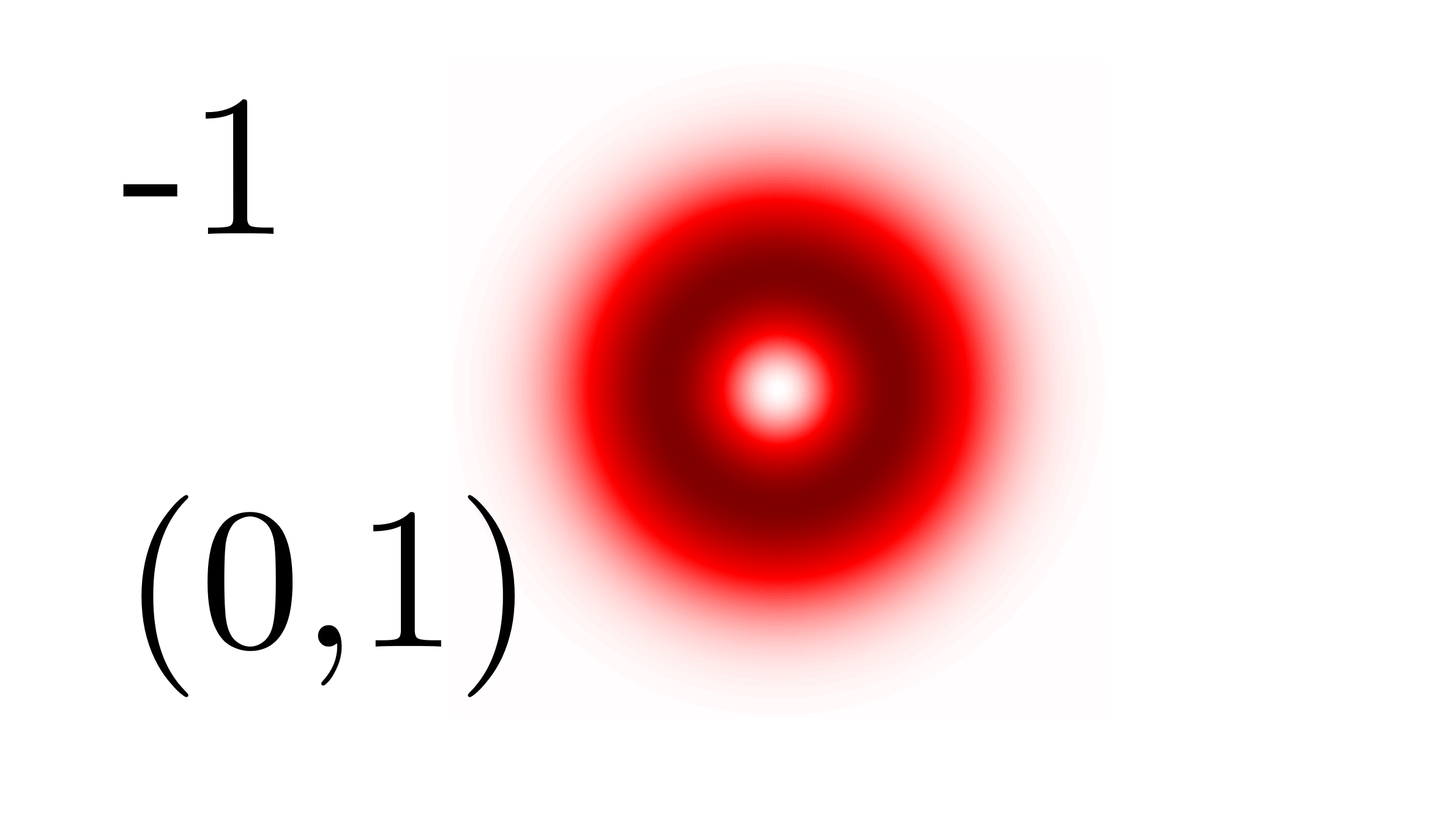} &\SetCell{bg=blue!20} $E_{\circlearrowleft}$ &\SetCell{bg=green!10} $\partial_\circlearrowright E_{\circlearrowleft}$ &\SetCell{bg=blue!20} $E_{\circlearrowright}$ \\
     \includegraphics[height=1cm]{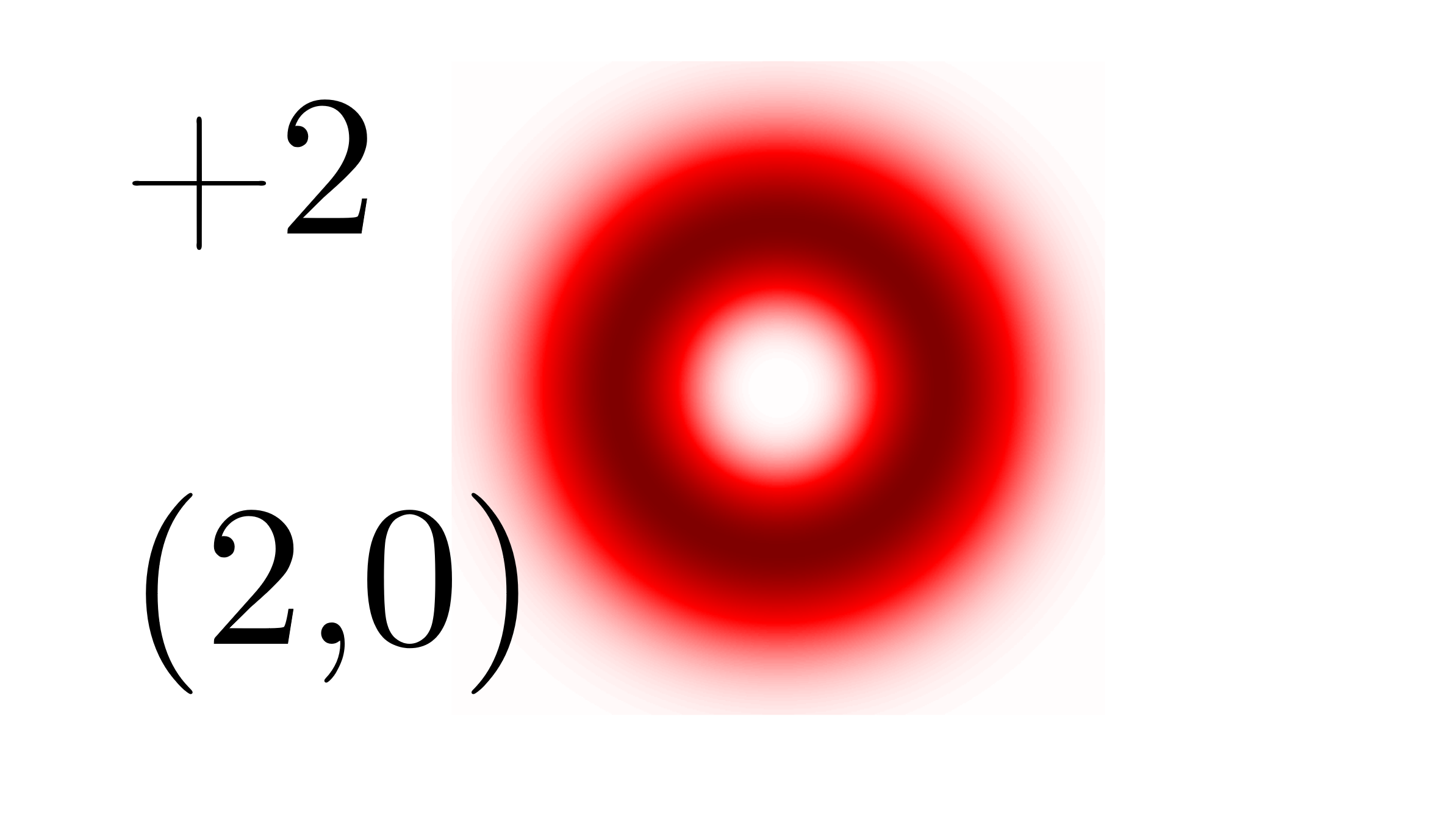} &\SetCell{bg=green!10} $\partial_\circlearrowright E_{\circlearrowleft}$ &\SetCell{bg=blue!20} $E_{\circlearrowright}$ &\SetCell{bg=green!10} $\partial_\circlearrowleft E_{\circlearrowright}$ \\
    \end{tblr}
    \label{tab:table2}
\end{table}

In order to further explore the mechanics behind these conservation/selection rules, we will exemplify three transitions in the rest of this letter. We start with the simplest transition: $|G\rangle \rightarrow |G\rangle$. The initial and final beams being the same, it corresponds to a $|i-j|=0^\text{th}$ order transition. Thus, only the first term of the multipolar development contributes i.e.: 
\begin{equation}
    g_{0,00\rightarrow 00}^{HG} = \frac{e}{\hbar v}\Phi \underbrace{\langle G \vert \mathds{1} \vert G\rangle}_{\delta_{0,0}\delta_{0,0}}
\end{equation}

\noindent where we omitted the dependence over $\textbf{R}_0, q_z$ and $\omega$ for brevity. From this equation, we immediately get: $ \Gamma^\text{PSEELS} = \vert g_{0,00\rightarrow 00}^{HG} \vert^2 = \frac{2 \pi e^2}{\hbar \omega} \tilde{\rho}_z (\textbf{R}_0,q_z,\omega)$. We retrieve thus the conventional EELS probability \eqref{classical_EELS}. This is expected since in an EELS experiment, the initial beam is roughly gaussian and with no post selection the outgoing beam will be a superposition of transverse shapes. The dominant term in probability will come from the Gaussian part of the outgoing beam. More generally, in any transition without transverse momentum exchange $\Delta n_x=\Delta n_y=0$, the transverse electromagnetic field must have no influence and the probability must only depend on $E_{z}$. These EELS-like transitions are highlighted by yellow cells in tables \ref{tab:table1} and \ref{tab:table2}.\\

We now move to the next order an examine $|i-j|=1^\text{st}$ order transition, for instance to a transition $|LG_{10}\rangle \rightarrow |G\rangle$, being the interaction of a left vortex beam with the target followed by the post-selection of a Gaussian wavefront. We readily obtain:
\begin{equation} \label{EQ:transition_order1_LG}
    g_{0, 10 \rightarrow 00}^{LG} =\frac{e w_0}{2\hbar v}E_{\circlearrowleft}\underset{\delta_{1-1,0}\delta_{0,0}}{\underbrace{\langle G \vert \hat{a}_\circlearrowleft \vert LG_{10}\rangle}}
\end{equation}

\noindent with $ E_{\circlearrowleft}$ the left circularly polarized electric field component. This transition corresponds to an electron state loosing a quantum $\Delta n_\circlearrowleft = -1$ of orbital angular momentum (OAM) and an optical field gaining a quantum of spin angular momentum (polarization) associated to the corresponding field amplitude $E_{\circlearrowleft}$. This effect - usually referred to as spin-orbit coupling - again illustrates the Noether theorem and was already observed in electron beams, e.g. \cite{vanacore_ultrafast_2019}. More generally, for arbitrary impinging HLG state, we retrieve the pEELS probability \eqref{pEELS} highlighted by blue cells in tables \ref{tab:table1} and \ref{tab:table2}. Thus, one can see that at order 1, only the dipolar term \eqref{Dipolar_moment_Hermite} contributes. A quantum field theory based treatment of this problem identified $\braket{\hat{\vec{d}}}$ to the polarization of the effective photon of the interaction - a quantity usually called optical polarization analogue (OPA, \cite{lourenco-martins_optical_2021,bourgeois_polarization-resolved_2022, bourgeois_optical_2023,nixon_inelastic_2024}). This shows that the electron beam then effectively behaves as a nanoscale source of polarized white light, thus enabling electron microscopes to push most of the polarization-selective optical experiment down to the deep sub-wavelength regime. 

So far, we have successfully recovered all the important results of EELS and pEELS of the literature, we shall now go further. Indeed, our model enables us to explore higher arbitrary orders without any further computational effort. We illustrate this on the next $|i-j|=2^\text{nd}$ order, in particular the transition $\vert HG_{11}\rangle \rightarrow \vert G \rangle$. We immediately get:
\begin{equation}
    g_{0, 11 \rightarrow 00}^{HG} =\frac{e w_0^2}{8 \hbar v}(\partial_x E_{y} + \partial_y E_{x})
\end{equation}

\noindent The right-hand side of the equation corresponds to the definition the local quadrupolar symmetry of a near-field shown on figure Fig.\ref{fig:Figure2}(b). Thus enabling to measure this quantity at a precise position on the sample and then to map it by scanning the electron beam over the sample as shown in \ref{fig:Figure2}(c,d).

Thinking back in terms of conservation law: during this quadrupolar transition the electron exchanges two quanta of linear momentum with the field $\Delta n_x=-1$ and $\Delta n_y=-1$. However, the electron still loses one quantum of energy $\hbar\omega$ i.e. one photon. Thus, one can see that for order $>1$ transitions - highlighted by green cells in tables \ref{tab:table1} and \ref{tab:table2} - the standard interpretation in terms of OPA employed in pEELS \cite{lourenco-martins_optical_2021,bourgeois_polarization-resolved_2022, bourgeois_optical_2023,nixon_inelastic_2024} fails: it is not possible to identify all the exchanged quanta to the sole polarization of an effective photon. This shows that for $k>1$, PSEELS goes further than mimicking optics and enters a regime unreachable with standard all-optical methods, becoming able to map so far inaccessible nano-optical quantities.

\begin{figure}
    \centering
    \includegraphics[width=\columnwidth]{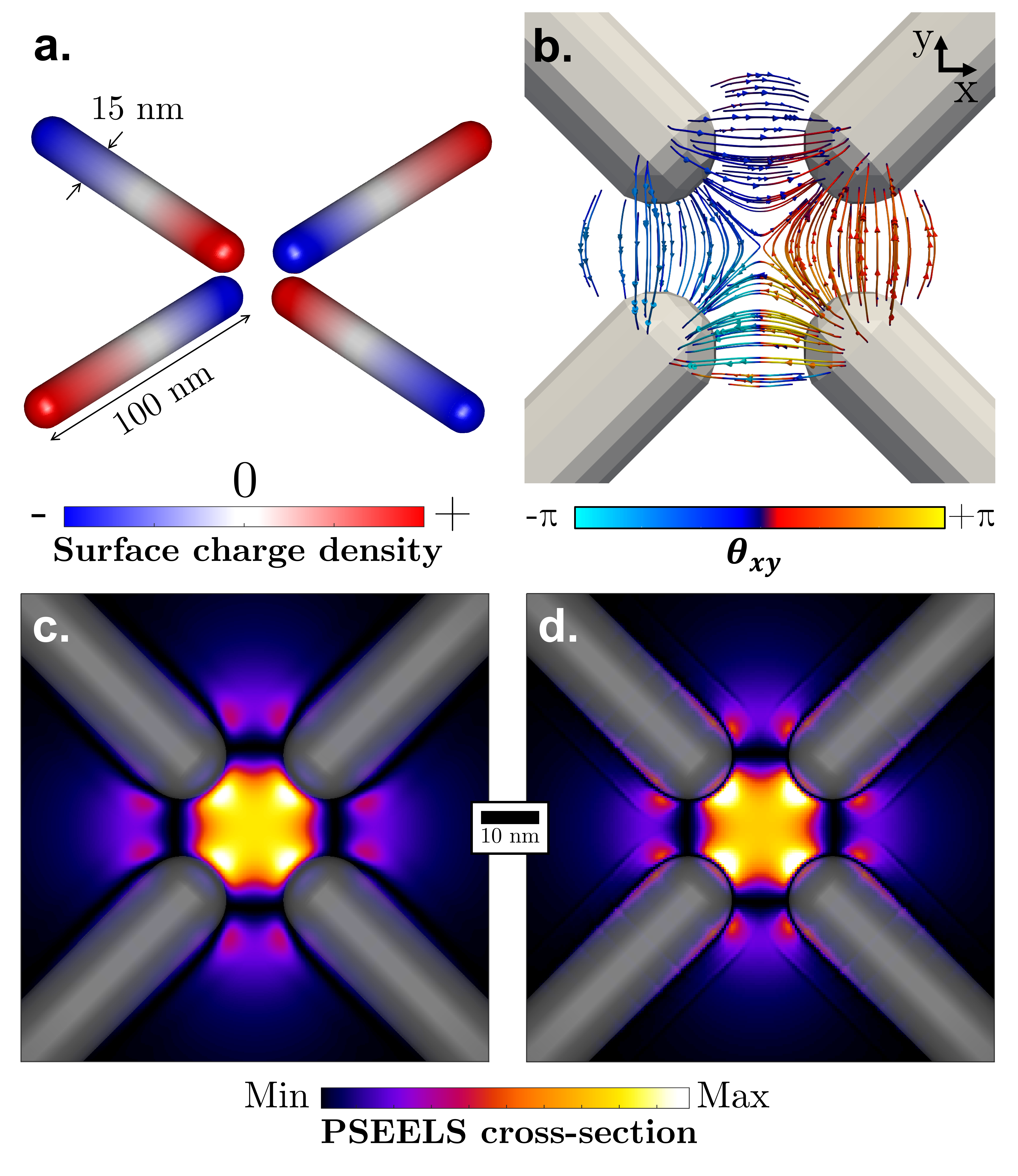}
    \caption{(a) 3D representation of four silver nanorods in-plane. (b) 3D representation of the electric field lines in the central gap, xy-plane top view with field direction arrows on the field lines. Colors of the field line representing the orientation angle $\theta_{xy}$ of the electric field projected on the xy-plane. (c,d) Spatially resolved maps across the gap : (c) PSEELS probability of transfer between a Hermite-Gauss $\vert HG_{1,1}\rangle$ and a Gaussian transverse state $\vert G\rangle$ for a 2nm wide electron beam ; (d) $|\partial_yE_x+\partial_xE_y|^2$, predicted to be proportional to the quantity measured in (c). Simulations done under MNPBEM library \cite{hohenester_mnpbem_2012}.}
    \label{fig:Figure2}
\end{figure}

Our formalism gives a simultaneous and intuitive treatment of all the possible transitions between Hermite-Gauss or Laguerre-Gauss states in PSEELS. It enables the calculation of the analytical transition amplitude between \emph{any} wavefronts as a sum of all the possible elementary HLG processes \eqref{Amplitude_HG_general} by expressing the wavefronts on the HG or LG basis. Conversely, it enables us to infer - without the need of heavy numerical computations - which beams to use to access any nano-optical quantities in PSEELS. Any optical quantity $C$ which can be expressed as a linear combination of the potential and its derivative $C(\{\phi,E,\partial_u E, \hdots,\partial^k_u E,\hdots \})$ is given by the exact same linear combination of HLG wavefronts associated with the corresponding transitions.

 In all of our developments the target degrees of freedom have been traced over $\sum_n \braket{n \vert . \vert 0}$. It should be kept in mind that EELS actually produces a sum of entangled target-electron states \cite{kfir_entanglements_2019,konecna_entangling_2022} e.g. for an initial electron-target state of the form $\ket{HG_{n_x,n_y},0}$, one gets:
\begin{equation}
    \ket{\psi_f}=\sum_{n_x',n_y',n} g_{0, n_x n_y \rightarrow n_x'n_y',n}^{HG}\ket{HG_{n'_x,n'_y},n}
\end{equation}

\noindent Thus, the selection of a specific transition $n_x n_y \rightarrow n_x'n_y'$ not only gives access to specific information in EELS encoded in $g_0$, but also post-selects the corresponding quantum state of the target. Applying the latest developments in electron-heralded photon state generation \cite{feist_cavity-mediated_2022,arend_electrons_2024}, the post-selection of specific free electron transitions (e.g. $|i-j|=2$) could enable the generation of quantum states of light with any desired symmetry (e.g. quadrupolar).
Eventually, common schemes in quantum optics harness the desexcitation of atoms with cavities to generate specific quantum state of light \cite{agarwal_nonclassical_1991}, albeit with parameters set by the intrinsic properties of the atom. Employing free electron orbitals instead of atomic orbitals would enable to bring flexibility to this method: here the nature of the "effective atom" is set by the electron beam shaper and sorter.
\\

\begin{acknowledgments}
This project has been funded in part by the French National Agency for Research under the program of future investment QUENOT (ANR-20-CE30-0033). The authors thank Mathieu Kociak and Jesse Groenen for illuminating discussions on the topic and for the help and support during the redaction of the manuscript.
\end{acknowledgments}

\PRLsep
\bibliographystyle{unsrt}
\bibliography{References.bib}

\appendix
\newpage
\onecolumngrid

\title{\textbf{Supplementary information for "Atomic-like selection rules in free electron scattering"}}
\author{Simon  Garrigou, Hugo  Lourenço-Martins}

\maketitle

\tableofcontents

\section{Phase shaped electron energy loss spectroscopy}

Accounting for the transverse profile changes of the electron beam through an energy loss interaction enriches the EELS experiment. The more complex the probe, the finer details we get. The goal of this section is to express properly the EELS interaction probability taking into account the transverse phase profile of the electron and its change during interaction. 

\subsection{EELS approximations}

In the EELS scheme, me consider a fast electron beam passing nearby a target (atom, molecule, dielectric or plasmonic structure...). The target presents a nano-localized optical field represented by a set of quantum states $\{ \ket{n}\}_{n\in\mathbb{N}}$. It can be excited by exchanging a virtual photon with the fast electron. As a consequence we describe the target as a polariton field \cite{basov_polaritons_2016,rivera_lightmatter_2020,kfir_optical_2021}, and generically denotes its states as \emph{photonic states}. To describe the system we apply a set of common approximations.\\

\textbf{Quasi-static}: By comparing the typical extension of the system's optical field $D$ to the wavelength of the interaction photon $\lambda = \frac{2 \pi c}{\omega}$, we can neglect the retardation effect, this is the \textbf{Quasi-static} approximation : 

\begin{equation}\label{quasi_static}
    \frac{\omega D}{c} \ll 1
\end{equation}
With $c$, the celerity of light and $\omega$ the angular frequency corresponding to the probed energy.

\textbf{Single scattering}: EELS scattering amplitudes are weak, it can be considered that the electron interacts once at maximum with the sample during its full propagation, this is the first Born approximation 
\begin{equation}\label{born_approx}
    \hat{U} \simeq 1 - \frac{i}{\hbar} \int \hat{V}(t) dt
\end{equation}
With $\hat{U}$ the system (electron + target) time evolution operator and $\hat{V}$ the Coulomb potential responsible for the interaction. $\hbar$ is the reduced Planck constant.

\textbf{Non-recoil}: In a transmission electron microscope (TEM) where these experiments are done, the typical speeds of the electron $v/c \sim 0.5 - 0.8$ enable us to neglect the lateral momentum transfers, this is the \textbf{non-recoil} approximation
\begin{equation}\label{non_recoil_approx}
    q_z = k_{f} - k_{i} \simeq \frac{\omega}{v}
\end{equation}
With $k_i$ and $k_f$ the z-component of the wavevector for respectively the initial and final state of the electron and $q_z$ the transferred momentum.

\textbf{Paraxial approximation} : TEM experiments involve small angles, justifying the use of the paraxial approximation \cite{lubk_chapter_2018} enabling to decouple the transverse and longitudinal part of the electron wavefunction
\begin{equation}\label{Paraxial_approx_posrep}
    \psi(\textbf{r}) \simeq \frac{1}{\sqrt{L}} e^{i k z} \Psi_\perp (\textbf{R})
\end{equation}
With $L$ the quantization length for the longitudinal wavefunction. In Dirac notation
\begin{equation}\label{Paraxial_approx}
    \vert \psi_i \rangle = \vert \textbf{k}_i^z \rangle \otimes \vert \Psi_{i,\perp} \rangle
\end{equation}

In this study we consider a \textbf{non diffracting} electron beam, the transverse wavefunction is unchanged during propagation.

\subsection{Electron description}
Following these approximations, the electron beam is described by a wavefunction separated in a transverse and longitudinal part as described by (\ref{Paraxial_approx}). In Dirac notation it is the outer product of a transverse state and a longitudinal state: under the different approximations these two subspaces do not mix during free propagation.
\begin{equation}
    \vert \psi_i \rangle = \vert k_i \rangle \otimes \vert \Psi_{i,\perp} \rangle
\end{equation}

\subsection{Interaction probability for general electron beam}
We consider the system composed of a target or photonic cavity and a free electron. EELS can be described as a quantum process of transition between two states of the system. With $ \vert i \rangle = \vert \psi_i,0 \rangle $ the initial state and $ \vert f \rangle = \vert \psi_f,n \rangle $ a final excited state. Respectively composed of the electron state $\vert \psi_i \rangle$ or $\vert \psi_f \rangle$  and the photonic states of the target $\vert 0 \rangle$ or $\vert n \rangle$ . We note the respective energies of these states: $\hbar\epsilon_i$, $\hbar\epsilon_f$, $\hbar\omega_0 $, and $ \hbar\omega_n$. From the Fermi's Golden rule, the EELS interaction rate reads 
\begin{equation}
    \frac{d \Gamma^\text{EELS}}{dt} = \frac{2\pi}{\hbar} \sum_{f} \Big| \langle f \vert \hat{V} \vert i \rangle \Big|^2 \delta(\hbar\epsilon_i - \hbar\epsilon_f +\hbar\omega_0 - \hbar\omega_n )
\end{equation}
Considering the propagation of the electron over a distance $L$ during a time $T$, the interaction probability reads
\begin{equation}
    \Gamma^\text{EELS} = \frac{2\pi T}{\hbar^2} \sum_{f} \Big| \langle f \vert \hat{V} \vert i \rangle \Big|^2 \delta(\epsilon_i - \epsilon_f +\omega_0 - \omega_n )
\end{equation}
With T the electron interaction time on the target. 

\begin{equation}
    \Gamma^\text{EELS} = \frac{2\pi T}{\hbar^2}\sum_{f}  \int d\omega \Big| \langle f \vert \hat{V} \vert i \rangle \Big|^2 \delta(\epsilon_i - \epsilon_f - \omega) \delta(\omega_0 - \omega_n - \omega)
\end{equation}
Initially the photonic state is empty of any excitation and the electron wavefunction is that of a fast electron in a microscope, eventually with a shaped transverse profile. The final state is obtained after a virtual photon emitted by the electron has populated a photonic state of the target. Hence during the process, the electron loses a quantum of energy, its transverse profile can change and a photonic state $\vert n \rangle$ for the target is populated by one excitation. The interaction leads to a superposition of final states corresponding to the different interaction paths. The interaction probability for each final state can be retrieved by sorting the outgoing electron beam and selecting the right state. This is done via  an electron spectrometer for energy resolution (fixed $\omega$) and post-selecting the transverse state with a state sorter for the electron beam.  
\begin{equation}
    \Gamma^\text{EELS} = \frac{2\pi T}{\hbar^2} \int d\omega \sum_{\psi_f,n} \Big| \langle \psi_f,n \vert \hat{V} \vert \psi_i,0 \rangle \Big|^2 \delta(\epsilon_i - \epsilon_f - \omega) \delta(\omega_0 - \omega_n - \omega)
\end{equation}
We evaluate the interaction for a specific energy. In consequence of the paraxial approximation we separate the transverse and longitudinal parts of the electron wavefunction. To simplify notations we adopt the convention $\vert a \rangle \otimes \vert b \rangle = \vert a \rangle\vert b \rangle = \vert a,b \rangle$
\begin{equation}
    \Gamma^\text{EELS}(\omega) = \frac{2\pi T}{\hbar^2} \sum_{k_f,\Psi_{f,\perp},n} \Big| \langle k_f \vert \langle \Psi_{f,\perp},n \vert \hat{V} \vert \Psi_{i,\perp},0 \rangle \vert k_i \rangle \Big|^2 \delta(\epsilon_i - \epsilon_f - \omega) \delta(\omega_0 - \omega_n - \omega)
\end{equation}
The longitudinal states for the free electron are expressed in a plane-wave basis. The sum can hence be made into an integral 
\begin{equation}
    \sum_{k_f} = \frac{L}{2\pi} \int dk_f
\end{equation}
Using the completeness relation $\mathds{1} = \int dz \vert z \rangle \langle z \vert$ and the longitudinal state position representation $\langle z \vert k_f \rangle = \frac{1}{\sqrt{L}} e^{i k_f z}$ 
\begin{equation}
\begin{split}
    \Gamma^\text{EELS}(\omega) = & \frac{L T}{\hbar^2} \sum_{n,\Psi_{f,\perp}} \int d k_f \left|\iint dz dz' \langle k_f \vert z \rangle \langle \Psi_{f,\perp},n \vert \langle z \vert \hat{V} \vert z \rangle \vert \Psi_{i,\perp},0 \rangle \langle z' \vert k_i \rangle \right|^2 \delta(\epsilon_i - \epsilon_f - \omega) \delta(\omega_0 - \omega_n - \omega)
\end{split}
\end{equation}

\begin{equation}
\begin{split}
    \Gamma^\text{EELS}(\omega) = & \frac{T}{\hbar^2 L} \sum_{n,\Psi_{f,\perp}} \int d k_f \left|  \langle \Psi_{f,\perp},n \vert \iint dz dz' e^{-i( k_f z - k_i z')}\langle z \vert \hat{V} \vert z \rangle \vert \Psi_{i,\perp},0 \rangle \right|^2 \delta(\epsilon_i - \epsilon_f - \omega) \delta(\omega_0 - \omega_n - \omega)
\end{split}
\end{equation}
In the narrow electron beam limit, we consider that the potential is local. $\langle z \vert \hat{V} \vert z' \rangle = \hat{V}(z,z') = \delta(z-z') \hat{V}(z)$. Hence

\begin{equation}
\begin{split}
    \Gamma^\text{EELS}(\omega) = & \frac{1}{\hbar^2 v} \sum_{n,\Psi_{f,\perp}} \int d k_f \left|  \langle \Psi_{f,\perp},n \vert \int dz e^{-i( k_f - k_i)z}\langle z \vert \hat{V} \vert z \rangle \vert \Psi_{i,\perp},0 \rangle \right|^2 \delta(\epsilon_i - \epsilon_f - \omega) \delta(\omega_0 - \omega_n - \omega)
\end{split}
\end{equation}
With $v=\frac{L}{T}$
\begin{equation}
\begin{split}
    \Gamma^\text{EELS}(\omega) = \frac{1}{\hbar^2 v} \sum_{n,\Psi_{f,\perp}} & \int d k_f \iint dz dz' e^{-i( k_f - k_i)(z-z')} \langle \Psi_{i,\perp},0 \vert \langle z' \vert \hat{V}^\dag \vert z' \rangle \vert \Psi_{f,\perp},n \rangle  \\ & \times \langle \Psi_{f,\perp},n \vert \langle z \vert \hat{V} \vert z \rangle \vert \Psi_{i,\perp},0 \rangle \delta(\epsilon_i - \epsilon_f - \omega) \delta(\omega_0 - \omega_n - \omega)
\end{split}
\end{equation}
With $q_z = k_f - k_i$ and $\epsilon_i - \epsilon_f = -q_z v$, we obtain
\begin{equation}
    \int dk_f e^{-i(k_f-k_i)(z-z')}  \delta(\epsilon_i - \epsilon_f - \omega) = \frac{1}{v}e^{i\frac{\omega}{v}(z-z')}
\end{equation}
The expression simplifies to
\begin{equation}
\begin{split}
    \Gamma^\text{EELS}(\omega) = & \frac{1}{\hbar^2 v^2} \sum_{n,\Psi_{f,\perp}}  \left|  \langle \Psi_{f,\perp},n \vert \int dz e^{i \frac{\omega}{v}z}\langle z \vert \hat{V} \vert z \rangle \vert \Psi_{i,\perp},0 \rangle \right|^2 \delta(\omega_0 - \omega_n - \omega)
\end{split}
\end{equation}
Where we identify the projected potential \cite{findlay_modeling_2008,dwyer_multislice_2005,coene_inelastic_1990,lubk_jacobs_2015}, z-Fourier transform of the interaction potential. 
\begin{equation}\label{eq_si:proj_potential_exp}
    \boxed{
    \hat{V}_\text{proj} (q_z,\omega) = \int dz \langle z \vert \hat{V} \vert z \rangle e^{i q_z z}
    }
\end{equation}

\begin{equation}
\begin{split}
    \Gamma^\text{EELS}(\omega) = & \frac{1}{\hbar^2 v^2} \sum_{n,\Psi_{f,\perp}}  \left|  \langle \Psi_{f,\perp} \vert \langle n \vert \hat{V}_\text{proj} \vert 0 \rangle \vert \Psi_{i,\perp} \rangle \right|^2 \delta(\omega_0 - \omega_n - \omega)
\end{split}
\end{equation}

The final electron beam is in a superposition of transverse profiles. Post-selecting the electron transverse state to a single state leads to

\begin{equation} \label{eq_si:EELS_qproba_fat}
\begin{split}
    \Gamma^\text{PSEELS}(\omega) = & \frac{1}{\hbar^2 v^2} \sum_{n}  \left|  \langle \Psi_{f,\perp} \vert \langle n \vert \hat{V}_\text{proj} \vert 0 \rangle \vert \Psi_{i,\perp} \rangle \right|^2 \delta(\omega_0 - \omega_n - \omega)
\end{split}
\end{equation}
 We now call this type of experiments Phase Shaped EELS (PSEELS). This formula contains the contribution of every photonic states at energy $\hbar \omega$. The non-zero contributions are those for which the transition moment for the photonic state matches the transition moment of the free electron transverse profile. This leads to a huge selectivity uppon proper choice of initial and final states for the electron. For brevity, we name $\omega_{0n} = \omega_0 - \omega_n$ and encompass the energy conservation in a new sum definition

 \begin{equation}
     \sum_n \delta(\omega_0 - \omega_n - \omega) = \sum_{\substack{n\\ \omega\ =\ \omega_{0n}}} = \tilde{\sum_n}
 \end{equation}
 
 \begin{equation} \label{eq_si:EELS_qproba}
 \boxed{
    \Gamma^\text{PSEELS}(\omega) =   \tilde{\sum_n}  \left| \frac{1}{\hbar v}  \langle \Psi_{f,\perp} \vert \langle n \vert \hat{V}_\text{proj} \vert 0 \rangle \vert \Psi_{i,\perp} \rangle \right|^2
}
\end{equation}

 \begin{equation} \label{eq_si:EELS_qproba_g0}
 \boxed{
    \Gamma^\text{PSEELS}(\omega) =   \tilde{\sum_n}  \left| \frac{1}{\hbar v}  \langle \Psi_{f,\perp} \vert \langle n \vert \hat{V}_\text{proj} \vert 0 \rangle \vert \Psi_{i,\perp} \rangle \right|^2 = \tilde{\sum_n} | g_{0,n} |^2 
}
\end{equation}
With $g_{0,n}$ the coupling constant. The next steps to study phase shaped interactions is to describe the transverse profiles of the electron beam and to express the projected potential. We will see in the following that the transverse profiles can be expressed through a ladder operator formalism and the projected potential can be made into a multipolar development. 

\section{Interaction Hamiltonian: Multipolar expansion}

\subsection{Projected potential: intuitive interpretation}
The EELS measurement is done after the electron propagation on a two dimensional detector, effectively summing the contributions along the z coordinate modulated by the electron phase. The interaction potential summed this ways leads to the projected potential, z-Fourier transform of the interaction potential, which emerges naturally from the EELS computation. It contains all the physical information of the interaction and accounts for the delocalized nature of the interaction. 

\subsection{Mutipolar expansion}
The electron beam waist $w_0$ (picometer) is small compared to the electrostatic potential variations scale (nanometer). Therefore the potential seen by the electron $\langle n \vert \hat{V}_\text{proj} \vert 0 \rangle = \hat{V}_\text{proj}(\textbf{R}_0,q_z,\omega)$ is considered constant over the waist of the electron beam, the same applies for its derivatives and thus the projected potential can be expressed via a multipolar development (equivalent to the Power-Zienau-Woolley form of the minimal-coupling) \cite{novotny_principles_2011}.
\begin{equation}\label{eq_si:projected_potential_mult_dev}
    \hat{V}_\text{proj} (q_z, \omega) = \underbrace{\hat{\rho} \hat{\Phi} (q_z, \omega)}_{\hat{V}_0} - \underbrace{\hat{\textbf{d}} \cdot \hat{\textbf{E}} (q_z, \omega)}_{\hat{V}_1} - \underbrace{\hat{\overleftrightarrow{Q}} \nabla \cdot \hat{\textbf{E}} (q_z, \omega)}_{\hat{V}_2} + ... + \text{Magnetic terms}
\end{equation}

Each term is composed of an operator acting on the electron transverse degree of freedom and an operator acting on the photonic degrees of freedom. If specified to a plasmonic structure, the photonic operator will lead to the electrostatic potential and its derivatives.

\subsection{Multipolar EELS}
The PSEELS coupling constant as seen in (\ref{eq_si:EELS_qproba_g0}) can be expressed for each photonic state $|n\rangle$ with the help of (\ref{eq_si:projected_potential_mult_dev}) as a multipolar EELS coupling constant. Each contribution will give informations on the physics of the system.

\begin{equation}\label{g_0_multipolar}
\begin{split}
    g_{0,n}(q_z,\omega)  & = \frac{1}{\hbar v} \langle \Psi_{f,\perp} \vert \langle n \vert  \hat{\rho}\hat{\Phi} - \hat{\textbf{d}} \cdot \hat{\textbf{E}} + ...\ \vert 0 \rangle \vert \Psi_{i,\perp} \rangle \\ & = g_{0,n}^{(0)} + g_{0,n}^{(1)} + g_{0,n}^{(2)} + ...
\end{split}
\end{equation}
As a consequence, The PSEELS probability (\ref{eq_si:EELS_qproba}) is also expressed as a multipolar probability

\begin{equation}\label{EELS_multipolar}
\begin{split}
    \Gamma^\text{PSEELS}(q_z,\omega)  & = \tilde{\sum_n}\Big|\frac{1}{\hbar v} \langle \Psi_{f,\perp} \vert \langle n \vert  \hat{\rho}\hat{\Phi} - \hat{\textbf{d}} \cdot \hat{\textbf{E}} + ...\ \vert 0 \rangle \vert \Psi_{i,\perp} \rangle\Big|^2 \\ & \cong \Gamma^{(0)} + \Gamma^{(1)} + \Gamma^{(2)} + ...
\end{split}
\end{equation}

Given the narrow beam condition needed for the multipolar development, the cross terms are either zero or orders of magnitude bellow the previous PSEELS order. We can neglect them to obtain this elegant order by order EELS expression.

\subsubsection{Gaussian order EELS}
We show here that the interaction due to the lowest order of the potential multipolar development is equivalent to classical EELS. We consider an initial and final Gaussian electron beam transverse profile, roughly equivalent to non shaped EELS. In a simplified vision, considering only the lowest order of the interaction 

\begin{equation}\label{eq_si:EELS_0}
    \Gamma^{(0)}(q_z,\omega) = \tilde{\sum_n}\Big|\frac{1}{\hbar v} \langle \Psi_{f,\perp} \vert \langle n \vert \hat{\rho} \hat{\Phi} \vert 0 \rangle \vert \Psi_{i,\perp} \rangle\Big|^2 
\end{equation}
The $\hat{\rho}$ operator acts on the transverse degree of freedom of the electron and the $\hat{\phi}$ acts on the photonic state only. They can be separated as follow
\begin{equation}\label{eq_si:EELS_0_separated}
    \Gamma^{(0)}(q_z,\omega) = \tilde{\sum_n}\Big|\frac{1}{\hbar v} \langle \Psi_{f,\perp}  \vert \hat{\rho} \vert \Psi_{i,\perp} \rangle \, \langle n \vert  \hat{\Phi} \vert 0 \rangle \Big|^2 
\end{equation}
We choose the initial and final transverse states as Gaussian states $\vert G \rangle$ which wavefunction has a waist $w_0$ and is centered at $\textbf{R}_0$.

\begin{equation}\label{}
    \Gamma^{(0)}(q_z,\omega) = \tilde{\sum_n}\left|\frac{e}{\hbar v}\int d\textbf{R} \Psi_{f,\perp}^*(\textbf{R}) \Psi_{i,\perp}(\textbf{R}) \langle n \vert \hat{\Phi}(\textbf{R}) \vert 0 \rangle  \right|^2 
\end{equation}

\begin{equation}
    \Psi_{f,\perp}(\textbf{R}) = \Psi_{i,\perp}(\textbf{R}) = \langle \textbf{R} \vert G \rangle = G(\textbf{R}) = \frac{1}{w_0}\sqrt{\frac{2}{\pi}} e^{-(\textbf{R}-\textbf{R}_0)^2/w_0^2}
\end{equation}
Under the narrow beam condition, where the typical length of variation for the potential is great compared to the beam waist, the transverse profile squared norm tends to a dirac delta function.

\begin{equation}
    \Big|G(\textbf{R})\Big|^2 \xrightarrow[w_0 \to 0]{} \delta(\textbf{R}-\textbf{R}_0) 
\end{equation}
Hence the electrostatic potential is probed at the beam position.
\begin{equation}\label{}
    \Gamma^{(0)}(q_z,\omega) = \tilde{\sum_n} \left| \frac{e}{\hbar v} \langle n \vert \hat{\Phi}(\textbf{R}_0) \vert 0 \rangle  \right|^2 
\end{equation}

This expression is in fact classical EELS, the following computations are aimed at retrieving the link to the local density of states (LDOS) as in \cite{losquin_link_2015}.

\begin{equation}
    \Gamma^{(0)}(q_z,\omega) = \frac{e^2}{\hbar^2 v^2}\tilde{\sum_n} \langle 0 \vert \hat{\Phi}^\dag(\textbf{R}_0) \vert n \rangle \langle n \vert \hat{\Phi}(\textbf{R}_0) \vert 0 \rangle 
\end{equation}
We "un-project" the potential
\begin{equation}
    \Gamma^{(0)}(q_z,\omega) = \frac{e^2}{\hbar^2 v^2}\iint dz dz' e^{i q_z (z'-z)} \tilde{\sum_n} \langle 0 \vert \hat{\Phi}^\dag(\textbf{R}_0,z) \vert n \rangle \langle n \vert \hat{\Phi}(\textbf{R}_0,z') \vert 0 \rangle 
\end{equation}
The electrostatic potential is hermitian, hence $\Phi^\dag = \Phi$. Using the formula for the electrostatic propagator in the framework of  Kubo's linear response theory \cite{bruus_manybody_2004}

\begin{equation}
    Im\left\{ W(r,r',\omega) \right\} = -\frac{\pi}{\hbar} \tilde{\sum_n} \langle 0 \vert \phi (\textbf{r}) \vert n \rangle \langle n \vert \phi(\textbf{r}') \vert 0 \rangle 
\end{equation}

\begin{equation}
    \Gamma^{(0)}(q_z,\omega) = \frac{e^2}{\hbar \pi v^2}\iint dz dz' e^{i q_z (z'-z)} Im\left\{ - W(\textbf{R}_0,\textbf{R}_0,z,z',\omega) \right\}
\end{equation}

$q_z = \frac{\omega}{v}$
\begin{equation}
    \Gamma^{(0)}(q_z,\omega) = \frac{e^2}{\hbar \pi \omega^2}\iint dz dz' q_z^2 e^{i q_z (z'-z)} Im\left\{ - W(\textbf{R}_0,\textbf{R}_0,z,z',\omega) \right\}
\end{equation}
By double integration by part along $z$ and $z'$.

\begin{equation}
    \Gamma^{(0)}(q_z,\omega) = \frac{e^2}{\hbar \pi \omega^2}\iint dz dz' e^{i q_z (z'-z)} Im\left\{ - \partial_z \partial_{z'} W(\textbf{R}_0,\textbf{R}_0,z,z',\omega) \right\}
\end{equation}
Leading to the identification of the Green dyad
\begin{equation}
    G_{zz} = \frac{1}{4\pi \omega^2} \partial_z \partial_{z'} W(\textbf{r},\textbf{r}')
\end{equation}

\begin{equation}
    \Gamma^{(0)}(q_z,\omega) = \frac{4 e^2}{\hbar}\iint dz dz' e^{i q_z (z'-z)} Im\left\{ - G_{zz}(\textbf{R}_0,\textbf{R}_0,z,z',\omega) \right\}
\end{equation}
Identifying the local density of states (LDOS) \cite{losquin_link_2015}
\begin{equation}
    \rho_{\boldsymbol{\mu}}(\textbf{r},\textbf{r},\omega) = - \frac{2 \omega}{\pi} Im\left\{ \boldsymbol{\mu}\cdot \overleftrightarrow{G}(\textbf{r},\textbf{r},\omega) \cdot\boldsymbol{\mu}  \right\}
\end{equation}

\begin{equation}
    \Gamma^{(0)}(\textbf{R}_0,q_z,\omega) = \frac{2 \pi e^2}{\hbar \omega}\iint dz dz' e^{i q_z (z'-z)} \rho_z (\textbf{R}_0,\textbf{R}_0,z,z',\omega)
\end{equation}
We retrieve the standard EELS probability being proportional to the LDOS.
\begin{equation}
    \Gamma^{(0)}(q_z,\omega) = \frac{2 \pi e^2}{\hbar \omega} \rho_z (\textbf{R}_0,\textbf{R}_0,q_z,-q_z,\omega)
\end{equation}
A transition from a gaussian state to a gaussian state is similar to the standard non shaped EELS experiment. The $\Gamma^{(0)}$ contribution is the leading order. Contributions from higher orders can be non-zero aswell but will be shown later to be orders of magnitude smaller. Indeed, standard EELS experiment consists of a gaussian incoming beam and no selection for the outgoing beam. The dominant term in probability will come from the Gaussian part of the outgoing beam.

\subsubsection{Field operators}
The field operators acting on the nano-optical field states give the expectation value $\langle n \vert \hat{A} \vert 0 \rangle$, (with $\hat{A}$ a field operator e.g $\hat{\textbf{E}}$ or $\hat{\Phi}$) proportional to the classical fields. For simpler notations we write

\begin{equation}\label{eq_si:field_operator_simpler}
    \langle n \vert \hat{A} \vert 0 \rangle = A_{n0} 
\end{equation}
For instance, if the target is specified to be a plasmonic structure, then these average values of the photonic operators will lead to the classical electrostatic potential and its derivatives. Indeed, while equation \eqref{eq_si:EELS_qproba_g0} involves a sum over all the possible quantum states of the target $\ket{n}$, it can be straightforwardly replaced by a sum over all the possible modes $m$ in the case of a classical system:
\begin{equation}\label{eq_si:quantum2classical}
    \sum_{\text{states } n} \vert g_{0,n} \vert^2 \xrightarrow[\text{classical field}]{} \sum_{\text{modes } m} \beta_m \; \vert g_{0,m}\vert^2
\end{equation}

\noindent where $g_{0,m}$ denotes the scattering amplitude per mode \cite{aguilar_selective_2023} and is a $\beta_m$ a scaling factor which represents the number of states per mode. In the widespread case of classical plasmonic field \cite{asenjo-garcia_dichroism_2014,ugarte_controlling_2015,guzzinati_probing_2017,zanfrognini_orbital_2019}, $\beta_m=4\hbar/\pi \;\Im\{-f_m\}$ with $f_m$ the modal spectral function \cite{aguilar_selective_2023}. Although the replacement rule \eqref{eq_si:quantum2classical} appears as an handy proxy ensuring a simple translation of our formalism between the quantum and classical regimes, it shall be kept in mind that modes and states are two completely different quantum optical concepts \cite{fabre_modes_2020} which should not be confused.\\

In later developments, we will focus on a single $n$ photonic mode of the system and omit the $n0$ subscript for clarity. PSEELS becomes
\begin{equation}\label{EELS_multipolar_simplified}
\begin{split}
    \Gamma^\text{PSEELS}(q_z,\omega)  & = \tilde{\sum_n}\Big|\frac{1}{\hbar v} \langle \Psi_{f,\perp} \vert  \hat{\rho}\Phi_{n0}(\textbf{R}_0,q_z,\omega) - \hat{\textbf{d}} \cdot \textbf{E}_{n0}(\textbf{R}_0,q_z,\omega) + ...\ \vert \Psi_{i,\perp} \rangle\Big|^2 \\ & \cong \Gamma^{(0)} + \Gamma^{(1)} + \Gamma^{(2)} + ...
\end{split}
\end{equation}

\section{Transverse electron wavefunction description}

We study transitions of the transverse electron wave-function between chosen states $\vert \Psi_{i,\perp} \rangle $ and $\vert \Psi_{f,\perp} \rangle$. Being a two dimensional problem, we can decompose every transverse wavefunction on the basis of states related to the two dimensional harmonic oscillator, inspired by \cite{nienhuis_paraxial_1993}. This brings in the ladder operator formalism from the standard textbook problem to express arbitrary wavefunction shape as a combination of creation and annihilation operators acting on a fundamental state. In a first approach, we pre- and post-select on the pure harmonic oscillator states to later generalise to any state. We here make a brief reminder on this topic.

\subsection{Harmonic oscillator reminders}
Considering a massive particle in a two dimensional space subjected to a two dimensional harmonic potential
\begin{equation}
    \hat{V} = \frac{1}{2} m\omega^2(\hat{x}^2+\hat{y}^2)
\end{equation}
The hamiltonian of the system is
\begin{equation}
    \hat{H} = \hat{H}_{xy} = \hat{H}_{x} + \hat{H}_{y} = \frac{\hat{p}_x^2+\hat{p}_y^2}{2 m} + \frac{1}{2} m\omega^2 (\hat{x}^2+\hat{y}^2)
\end{equation}

With $\hat{H}_{x}$ and $\hat{H}_{y}$ the 1D harmonic oscillator hamiltonian. To prepare for the electron description we use the notation
\begin{equation}
    \frac{\sqrt{2}}{w_0} = \sqrt{\frac{m \omega}{\hbar}}
\end{equation}
Where $w_0$ corresponds to the waist of the electron beam in the PSEELS study. The eigenvalues $\vert \chi \rangle$ of the hamiltonian are found using the time independent Schrödinger equation
\begin{equation}
    \hat{H} \vert \chi \rangle =  E\vert \chi \rangle
\end{equation}
The x and y coordinates remain independent, thus the energy spectrum is the sum of two one dimension harmonic oscillator.
\begin{equation}
    E = \left( n_x +\frac{1}{2} \right) \hbar \omega +\left( n_y +\frac{1}{2} \right) \hbar \omega = \left( n_x +n_y + 1 \right) \hbar \omega 
\end{equation}
The eigenvalues of the hamiltonian are degenerate. Two different sets of operator are used to properly find and index the eigenstates : the "cartesian" $\{\hat{H}_{xy}, \hat{H}_{x}, \hat{H}_{y}\}$ or "angular momentum" $\{\hat{H}_{xy},\hat{L}_{z}\}$ where $\hat{L}_{z}$ is the orbital angular momentum operator. Consequence of this fact is the existence of two complete sets of orthogonal eigenstates describing the system, the Hermite-Gauss $\vert HG_{n_x,n_y}\rangle$ for the "cartesian" and the circular $\vert LG_{n_\circlearrowright,n_\circlearrowleft} \rangle$ for the angular momentum. The eigenvalue of $\hat{H}_{xy}$ is the number of quanta of vibration in the system: $n_x+n_y = n_\circlearrowleft + n_\circlearrowright$ ; The eigenvalue of $\hat{L}_{z}$ is $n_\circlearrowleft - n_\circlearrowright$, balance of left and right circular vibration quanta. The canonical way to study this system is to define the ladder operators in the Cartesian basis, creating or annihilating one quantum of vibration in the corresponding direction. In this system, acting on the x direction has no impact on the y direction.

\begin{equation}\label{a_x}
    \hat{a}_x^{} = (\frac{1}{w_0} \hat{x} + i \frac{w_0 \hat{p_x}}{ 2\hbar})
\end{equation}

\begin{equation}\label{a_y}
    \hat{a}_y^{} = (\frac{1}{w_0} \hat{y} + i \frac{w_0 \hat{p_y}}{ 2\hbar})
\end{equation}

The ladder operators obey the following commutation relation
\begin{equation}
    [\hat{a}_i,\hat{a}_j^\dag] = \delta^{}_{i,j} \mathds{1}
\end{equation}

The system's hamiltonian can be expressed in term of these operators as : 
\begin{equation}
    \hat{H} = \hbar\omega ( 1 + \hat{N}_x + \hat{N}_y )
\end{equation}
With the number operator
\begin{equation}
    \hat{N}_i = \hat{a}_i^\dag \hat{a}_i 
\end{equation}

\subsection{Ladder operator to build wavefunctions}

Eigenstates of the system can be built from the fundamental eigenstate $\vert 0 \rangle = \vert G \rangle$ which is the 2D gaussian state \cite{claude_cohen-tannoudji_cohen-tannoudji_2019}. Two sets of eigenstates can be used to describe properly the system, the Hermite-Gauss states for cartesian coordinates and the circular states for the "angular momentum" coordinates, the later can be mapped to Laguerre-Gauss states. Eigenstates in a given basis can be indexed by the number of each ladder operators needed to generate it starting from the fundamental gaussian mode
\begin{equation}
   \langle \textbf{r} \vert G \rangle = G(x,y) = \frac{1}{w_0}\sqrt{\frac{2}{\pi}} e^{- (x^2+y^2)/w_0^2}
\end{equation}
Which is normalized so that $\langle G \vert G \rangle = 1$.

\subsubsection{Cartesian : Hermite-Gauss}
The first set of operator sharing a basis of eigenvectors is $\{\hat{H}_{xy},\hat{H}_{x},\hat{H}_{y}\}$. We name it the cartesian coordinate basis. Eigenstates are the Hermite-Gauss states $\vert HG_{n_x,n_y} \rangle$. the cartesian ladder operators act on these states following
\begin{equation}\label{HG_ladder_rules}
\begin{split}
    \hat{a}_x \vert HG_{n_x,n_y} \rangle &= \sqrt{n_x}\ \vert HG_{n_x-1,n_y} \rangle \\
    \hat{a}_x^\dag \vert HG_{n_x,n_y} \rangle &= \sqrt{n_x+1}\ \vert HG_{n_x+1,n_y} \rangle\\
    \hat{a}_y \vert HG_{n_x,n_y} \rangle &= \sqrt{n_y}\ \vert HG_{n_x,n_y-1} \rangle \\
    \hat{a}_y^\dag \vert HG_{n_x,n_y} \rangle &= \sqrt{n_y+1}\ \vert HG_{n_x,n_y+1} \rangle
\end{split}
\end{equation}
Every HG state hence can be expressed from the Gaussian fundamental state
\begin{equation}\label{eq_si:HG_ladder_op}
    \vert HG_{n_x,n_y}^{} \rangle = \frac{1}{\sqrt{n_x! n_y!}} (\hat{a}^\dag_x)^{n_x} (\hat{a}^\dag_y)^{n_y} \vert G \rangle 
\end{equation}

Expressed as a wavefunction
\begin{equation}\label{Hermite-Gauss expression}
   \langle \textbf{r} \vert HG_{n_x,n_y} \rangle = HG_{n_x,n_y}(x,y) = \frac{H_{n_x}\left(\frac{x\sqrt{2}}{w_0} \right) H_{n_y}\left(\frac{y\sqrt{2}}{w_0} \right)}{w_0\sqrt{2^{n_x+n_y -1} \pi n_x! n_y!}} e^{- (x^2+y^2)/w_0^2} 
\end{equation}
With $H_{n_x}(x)$ the Hermite polynomial of degree $n_x$. We here define the degree of the transverse electron wavefunction as the number of creation operators used to build it from the Gaussian state $n_x+n_y$. It is also the sum of the degrees of the Hermite polynomials describing it. Given this formula we can express every modes of this basis and we have a method to relate them to the fundamental gaussian state. 

\subsubsection{Circular : Laguerre-Gauss}
The second set of operator sharing a basis of eigenvectors is $\{\hat{H}_{xy},\hat{L}_z\}$. The eigenstates have a well defined angular momentum. They can be mapped to the Laguerre Gauss modes. The orbital angular momentum operator reads: 

\begin{equation}
    \hat{\textbf{L}} = \hat{\textbf{R}} \times \hat{\textbf{P}}
\end{equation}
With $\hat{\textbf{R}} $ and $ \hat{\textbf{P}}$ the position and momentum operators respectively. Projected on the z-axis:
\begin{equation}
    \hat{L}_z = \hat{x}\hat{p}_y - \hat{y}\hat{p}_x
\end{equation}
Also expressed in the position representation (($r,\phi$) coordinates) as
\begin{equation}
    \langle \textbf{r} \vert \hat{L}_z = \frac{\hbar}{i} \frac{\partial}{\partial \phi} \langle \textbf{r} \vert
\end{equation}
The z-orbital angular momentum operator can be expressed using cartesian ladder operators
\begin{equation}
    \hat{L}_z = i \hbar (\hat{a}_x^{} \hat{a}_y^\dag - \hat{a}_x^\dag \hat{a}_y^{} )
\end{equation}

We introduce $a_\circlearrowright^\dag$ and $a_\circlearrowleft^\dag$, that can be seen as "circular creation operators" \cite{claude_cohen-tannoudji_cohen-tannoudji_2019}. Each one creates a quantum of circular vibration, that is, carrying one unit of angular momentum. The conjugate operator anihilates the corresponding quatum. We take the reversed convention of \cite{claude_cohen-tannoudji_cohen-tannoudji_2019} as we define $\hat{a}_\circlearrowleft^{\dag}$ as the operator adding a positive quantum of orbital angular momentum to the system, e.g $\hat{L}_z\ \hat{a}_\circlearrowleft^\dag \vert G \rangle = +\hbar\ \hat{a}_\circlearrowleft^\dag \vert G \rangle$

  \begin{equation}\label{left a/c op}
\begin{split}
    \hat{a}_\circlearrowleft^\dag = \frac{1}{\sqrt{2}} (\hat{a}_x^\dag + i \hat{a}_y^\dag) \\
    \hat{a}_\circlearrowleft^{}= \frac{1}{\sqrt{2}} (\hat{a}_x^{} - i \hat{a}_y^{} ) 
\end{split}
\end{equation}

\begin{equation}\label{right a/c op}
\begin{split}
    \hat{a}_\circlearrowright^\dag = \frac{1}{\sqrt{2}} (\hat{a}_x^\dag - i \hat{a}_y^\dag) \\
    \hat{a}_\circlearrowright^{}= \frac{1}{\sqrt{2}} (\hat{a}_x^{} + i \hat{a}_y^{} ) 
\end{split}
\end{equation}

With these we can rewrite the operators

\begin{equation}
    \hat{H}_{xy} = ( \hat{a}_\circlearrowleft^\dag \hat{a}_\circlearrowleft^{}+ \hat{a}_\circlearrowright^\dag \hat{a}_\circlearrowright^{}+ 1 ) \hbar \omega = ( \hat{N}_\circlearrowleft + \hat{N}_\circlearrowright + 1 ) \hbar \omega
\end{equation}

\begin{equation}
    \hat{L}_z = ( \hat{a}_\circlearrowleft^\dag \hat{a}_\circlearrowleft^{} - \hat{a}_\circlearrowright^\dag \hat{a}_\circlearrowright^{} ) \hbar = ( \hat{N}_\circlearrowleft - \hat{N}_\circlearrowright ) \hbar
\end{equation}

$\hat{H}_{xy}$ appears as an undistinguished count of the circular quanta in the system whereas $\hat{L}_z$ measures the balance of left and right quanta in the system. Under the present convention, a left rotating object carries a positive angular momentum. Acting on the Gaussian state of the system with these ladder operators gives new states with well-defined angular momenta as are the Laguerre-Gauss modes. We name them $\vert LG \rangle$, however the position of the indices brings a nuance.
\begin{equation}\label{eq_si:LG_ladder_op}
    \vert LG_{n_\circlearrowright,n_\circlearrowleft}\rangle = \frac{(\hat{a}_\circlearrowright^\dag)^{n_\circlearrowright}(\hat{a}_\circlearrowleft^\dag)^{n_\circlearrowleft}}{\sqrt{n_\circlearrowright! n_\circlearrowleft!}} \vert G \rangle 
\end{equation}
Mapping to Laguerre-Gauss states is particularly interesting : they form a basis commonly used in the description of optical vortex beams and have a lot of useful representation tools. It can be shown that \cite{karimi_radial_2014,claude_cohen-tannoudji_cohen-tannoudji_2019,beijersbergen_astigmatic_1993}
\begin{equation}
    \vert LG_{n_\circlearrowright,n_\circlearrowleft}\rangle = (-1)^{p} \vert LG_p^l \rangle
\end{equation}
With $l = (n_\circlearrowleft - n_\circlearrowright) \in \mathbb{Z}$, the orbital angular momentum quantum number and $p = \frac{1}{2}(n_\circlearrowleft + n_\circlearrowright - |n_\circlearrowleft - n_\circlearrowright|) =min(n_\circlearrowright,n_\circlearrowleft) \in \mathbb{N}$ the radial quantum number. 

The position representations of Laguerre-Gauss states are
\begin{equation}
    \langle \textbf{r} \vert LG_{p}^l \rangle = LG_p^l (r,\varphi) = \sqrt{\frac{2 p!}{\pi (p + |l|)!}}\frac{1}{w_0} \left(\frac{\sqrt{2}r}{w_0} \right)^{|l|}  L_p^{|l|}\left(\frac{2 r^2}{w_0^2}\right)\ e^{-r^2/w_0^2} e^{i l \varphi}
\end{equation}

The wavefunction is expressed in term of the generalized Laguerre polynomials $L_p^{|l|}(r)$. Note the subtle difference between these two family of states. The $\vert LG_{n_\circlearrowright,n_\circlearrowleft}\rangle$ straightforwardly expresses the wavefunction in term of number of left or right quanta and the Laguerre-Gauss is easier to represent since more commonly used, even if the $p$ quantum number is harder to interpret. These two sets of functions are identical to a $(-1)^p$ factor, this nuance is not exposed in the main text for brevity. The following table represents the correspondence of wavefunctions $\vert LG \rangle$ indexed by $n_\circlearrowright$ and $n_\circlearrowleft$ and $\vert LG \rangle$ indexed by $p$ and $l$ and represented as $(p,l)$ in the table
\begin{center}
\begin{tabular}{ |c|c|c|c|c|c|c| } 
 \hline
 \diagbox[width=5.5em, height=3\line]{$n_\circlearrowleft$}{$n_\circlearrowright$} & 0 & 1 & 2 & 3 & 4 & $n_\circlearrowright$ \\ 
  \hline
 0 & (0,0) & (0-1) & (0,-2) & (0,-3) & (0,-4) &\\ 
  \hline
 1 &(0,1) & \textbf{\textit{(1,0)}} & \textbf{\textit{(1,-1)}} & \textbf{\textit{(1,-2)}} &  &\\ 
 \hline
  2 & (0,2) & \textbf{\textit{(1,1)}} & (2,0) & (2,-1) & & \\ 
 \hline
  3 & (0,3) & \textbf{\textit{(1,2)}} & (2,1) & \textbf{\textit{(3,0)}} & & \\ 
 \hline
  4 & (0,4) & & & & &\\ 
  \hline
  $n_\circlearrowleft$ & & & & & & $p=min(n_\circlearrowright,n_\circlearrowleft)$ \\
  &&&&&& $l=n_\circlearrowleft-n_\circlearrowright$\\
 \hline
\end{tabular}
\end{center}

Where $\textbf{\textit{(p,l)}}$ functions have a global minus sign compared to their $n_\circlearrowright,n_\circlearrowleft$ equivalent. Creating a left \emph{and} a right quanta of excitation on the gaussian state leads to raising the $p$ values, hence adding a radial node to the to the wavefunction without changing the orbital angular momentum ("dark ring" in the intensity profile). The ladder operator rules are

\begin{equation}\label{LG_ladder_rules}
    \begin{split}
    \hat{a}_{\circlearrowright}^{}\vert LG_{n_\circlearrowright,n_\circlearrowleft} \rangle &= \sqrt{n_\circlearrowright} \vert LG_{n_\circlearrowright-1,n_\circlearrowleft} \rangle \\
    \hat{a}_{\circlearrowleft}^{}\vert LG_{n_\circlearrowright,n_\circlearrowleft} \rangle &= \sqrt{n_\circlearrowleft} \vert LG_{n_\circlearrowright,n_\circlearrowleft-1} \rangle \\
    \hat{a}_{\circlearrowright}^\dag\vert LG_{n_\circlearrowright,n_\circlearrowleft} \rangle &= \sqrt{n_\circlearrowright+1} \vert LG_{n_\circlearrowright+1,n_\circlearrowleft} \rangle \\
    \hat{a}_{\circlearrowleft}^\dag\vert LG_{n_\circlearrowright,n_\circlearrowleft} \rangle &= \sqrt{n_\circlearrowleft+1} \vert LG_{n_\circlearrowright,n_\circlearrowleft+1} \rangle \\
\end{split}
\end{equation}

\subsection{Basis interconnectivity}

There is a direct mapping of LG to HG, discussed for instance in these articles \cite{allen_orbital_1992,abramochkin_beam_1991,beijersbergen_astigmatic_1993,nienhuis_paraxial_1993}. We adapt here the notations to our formalism. Defining $N = n_x + n_y = n_\circlearrowright + n_\circlearrowleft$, the order of the transverse profile, and the usual $p=min(n_\circlearrowright , n_\circlearrowleft)$ for LG
\begin{equation}
    \vert LG_{n_\circlearrowleft,n_\circlearrowright}\rangle = (-1)^p\sum^N_{k=0} i^k b(n_\circlearrowright,n_\circlearrowleft,k) \vert HG_{N-k,k}\rangle
\end{equation}
With
\begin{equation}
    b(n_\circlearrowright,n_\circlearrowleft,k) = \left( \frac{(N-k)!k!}{2^N n_\circlearrowright! n_\circlearrowleft!}\right)^{1/2} \frac{1}{k!}\frac{d^k}{dt^k}\left[ (1-t)^{n_\circlearrowright} (1+t)^{n_\circlearrowleft} \right]_{t=0}
\end{equation}

This relations shows that an order $N$ Laguerre-Gauss beam is expressed exclusively as a sum of order $N$ Hermite-Gauss beams. Meaning, in some sense, that there is a conserved quantity in the decomposition and every order is well separated. Interestingly there exists nearly the same relation linking HG along x,y axis and x',y' axis: tilted by 45°. It is Identical as for the LG but in the real space (no imaginary term):
\begin{equation}
    HG_{n,m}\left(\frac{x+y}{\sqrt{2}},\frac{x-y}{\sqrt{2}} \right) = \sum^N_{k=0} b(n,m,k) HG_{N-k,k}(x,y)
\end{equation}
We can rewrite the HG to LG relation under the previously defined operator algebra for 2D functions.

\begin{equation}
    \left( \frac{(\hat{a}^\dag_\circlearrowleft)^{n_\circlearrowleft}(\hat{a}^\dag_\circlearrowright)^{n_\circlearrowright}}{\sqrt{n_\circlearrowleft!n_\circlearrowright!}} \right) \vert G\rangle = (-1)^{p} \left[\sum^N_{k=0} i^k b(n_\circlearrowleft,n_\circlearrowright,k) \frac{(\hat{a}^\dag_x)^{(N-k)}(\hat{a}^\dag_y)^k}{\sqrt{(N-k)!k!}} \right] \vert G\rangle
\end{equation}

Which simplifies by expanding the $b$ term as: 

\begin{equation}
    (\hat{a}^\dag_\circlearrowleft)^{n_\circlearrowleft}(\hat{a}^\dag_\circlearrowright)^{n_\circlearrowright} \vert G\rangle = \frac{(-1)^{p}}{\sqrt{2^N}}\left[\sum^N_{k=0}  \frac{i^k}{k!}\frac{d^k}{dt^k} [(1-t)^{n_\circlearrowright} (1+t)^{n_\circlearrowleft}]_{t=0} (\hat{a}^\dag_x)^{(N-k)}(\hat{a}^\dag_y)^k \right] \vert G\rangle   
\end{equation}
Hence we have the operator relation linking a product of creation operators in the LG basis to a product of creation operators in the HG basis.
\begin{equation}
    (\hat{a}^\dag_\circlearrowleft)^{n_\circlearrowleft}(\hat{a}^\dag_\circlearrowright)^{n_\circlearrowright} = \frac{(-1)^{min(n_\circlearrowleft,n_\circlearrowright)}}{\sqrt{2^N}}\left[\sum^N_{k=0}  \frac{i^k}{k!}\frac{d^k}{dt^k} [(1-t)^{n_\circlearrowright} (1+t)^{n_\circlearrowleft}]_{t=0} (\hat{a}^\dag_x)^{(N-k)}(\hat{a}^\dag_y)^k \right]
\end{equation}
At this point, electron transverse wavefunctions can be expressed on two different basis : The Hermite Gauss $\vert HG_{n_x,n_y} \rangle$ and the Laguerre-Gauss $\vert LG_{n_\circlearrowright,n_\circlearrowleft} \rangle$. Each element of these basis can be expressed as ladder operators acting on a gaussian state. The decomposition of Laguerre Gauss modes on Hermite-Gauss basis is a useful tool to express PSEELS experiments where the transition occurs between two states from different basis, for instance $\vert \Psi_{i,\perp} \rangle = \vert LG_{1,2} \rangle$, $\vert \Psi_{f,\perp} \rangle = \vert HG_{1,1} \rangle$.

\subsection{General transverse wavefunction expression}
Any transverse electron beam can be expressed as a combination of $\vert HG_{n_x,n_y}\rangle$ or $\vert LG_{n_\circlearrowleft,n_\circlearrowright}\rangle$.

\begin{equation}
    \begin{split}
        \vert \Psi_\perp \rangle & = \sum_{n_x,n_y} c_{n_x,n_y} \vert HG_{n_x,n_y}\rangle \\
        \vert \Psi_\perp \rangle & = \sum_{n_\circlearrowleft,n_\circlearrowright} c_{n_\circlearrowleft,n_\circlearrowright} \vert LG_{n_\circlearrowleft,n_\circlearrowright}\rangle
    \end{split}
\end{equation}
With 
\begin{equation}
    \begin{split}
        c_{n_x,n_y} &= \langle HG_{n_x,n_y} \vert \Psi_\perp \rangle \\
        c_{n_\circlearrowleft,n_\circlearrowright} &= \langle LG_{n_\circlearrowleft,n_\circlearrowright} \vert \Psi_\perp \rangle
    \end{split}
\end{equation}
Which in turn can be expressed through ladder operators (\ref{eq_si:HG_ladder_op},\ref{eq_si:LG_ladder_op}). In the following we will only study pure states (Laguerre- or Hermite- Gauss), generalization is then straightforward.

\subsection{Order of a pure transverse state}
The order $i$ of a pure HG or LG transverse state is defined as the number of creation operators required to build it from the gaussian state. We have $i = n_x+n_y$ or $i=n_\circlearrowleft + n_\circlearrowright$. 

\subsection{Transition order between pure states}
 Any pure transverse state can be transformed into another pure transverse state by application of a product of ladder operators. Then, the order $|i-j|$ of the transition between two transverse states $i$ and $j$ is defined as the number of operators needed to transit from $i$ to $j$. A first order transition corresponds to $|HG_{1,0}\rangle \rightarrow |G\rangle$ or $|LG_{1,0}\rangle \rightarrow |LG_{1,1}\rangle$ for instance.\\
 Rigorously, the order of the transition $|HG_{n_x,n_y}\rangle \rightarrow |HG_{n_{x'},n_{y'}}\rangle$ is defined as 
 \begin{equation}
     |i-j| = |n_{x'}-n_{x}| + |n_{y'}-n_{y}| = |\Delta n_x| + |\Delta n_y|
 \end{equation}
This subtlety in the definition of the transition order has an impact for instance on transition $|HG_{1,0}\rangle \rightarrow |HG_{0,1}\rangle$ where the simple difference $(i-j) = 0$ but the transition order $|i-j| = 2$. The definition is easily generalized to LG states.

\section{Multipolar interaction potential under ladder operator formalism}
\subsection{Interaction potential with operator algebra}
Any transverse electron beam profile can be expressed via a ladder operator combination, the same is done for the projected potential multipolar development (\ref{eq_si:projected_potential_mult_dev}) at every orders. Given the expressions for the ladder operators in the two coordinate systems (\ref{a_x}),(\ref{a_y}), (\ref{left a/c op}) and (\ref{right a/c op}), the position operators can be re-expressed as

\begin{equation}\label{x_expression_circlearrowleftadder_op_}
        \hat{x} = \frac{w_0}{2} \left( \hat{a}_x^{} + \hat{a}_x^\dag \right) = \frac{w_0}{2\sqrt{2}} \left( \hat{a}_\circlearrowright^{} + \hat{a}_\circlearrowleft^{} + \hat{a}_\circlearrowright^\dag + \hat{a}_\circlearrowleft^\dag \right)
\end{equation}
\begin{equation}\label{y_expression_circlearrowleftadder_op_}
    \hat{y} = \frac{w_0}{2} \left( \hat{a}_y^{} + \hat{a}_y^\dag \right) = \frac{i w_0}{2\sqrt{2}} \left( \hat{a}_\circlearrowleft^{} - \hat{a}_\circlearrowright^{} - \hat{a}_\circlearrowleft^\dag + \hat{a}_\circlearrowright^\dag \right)
\end{equation}
As a consequence, each term $\hat{V}_k$ of the multipolar development can be expressed as a combination of ladder operators. Each term only involves products of $k$ ladder operators $\hat{a}^{(k)}\equiv(\hat{a}_u \times \hdots \times \hat{a}_{u'})_{k \text{ factors}}$
\begin{equation}
     \braket{n\vert\hat{V}_\text{proj}\vert 0}=\sum_{k} \braket{n\vert\hat{V}_k[\hat{a}^{(k)}]\vert 0} = \sum_{k} \hat{V}_{n0,k}[\hat{a}^{(k)}]
\end{equation}
The integer $k$ will be called the order of the transition in the following.

\subsection{Dipolar term}

The dipolar term in the potential is expressed as

\begin{equation}
    \hat{V}_1 = - \hat{\textbf{d}} \cdot \textbf{E}_{n0} (\textbf{R}_0,q_z, \omega) = e \left( \hat{x} E_{n0,x} (\textbf{R}_0,q_z, \omega) + \hat{y} E_{n0,y} (\textbf{R}_0,q_z, \omega) \right) 
\end{equation}
From now on, the $(\textbf{R}_0,q_z, \omega)$ dependence of the field quantities and the $n0$ subscript on the components of the electric field will be implicit for readability purpose. We express it in both basis of the harmonic oscillator states. 
\subsubsection{Dipolar potential : Hermite-Gauss}

\begin{equation}\label{Dipolar_Hermite}
    \hat{V}_1 = \frac{e w_0}{2} \left[ (\hat{a}_x^{} + \hat{a}_x^\dag) E_x +  (\hat{a}_y^{} + \hat{a}_y^\dag) E_y \right]
\end{equation}

\subsubsection{Dipolar potential : Laguerre-Gauss}
\begin{equation}\label{Dipolar_Laguerre}
    \hat{V}_1 = \frac{e w_0}{2} \left[ (\hat{a}_\circlearrowleft^{} + \hat{a}_\circlearrowright^\dag) E_\circlearrowleft  +  (\hat{a}_\circlearrowright^{} + \hat{a}_\circlearrowleft^\dag) E_\circlearrowright  \right]
\end{equation}
Where we name 
\begin{equation}\label{E_circlearrowleft_def}
    E_\circlearrowleft = \frac{1}{\sqrt{2}} (E_x + i E_y)
\end{equation}
and 
\begin{equation}\label{E_circlearrowright_def}
    E_\circlearrowright = \frac{1}{\sqrt{2}} (E_x - i E_y) 
\end{equation}
The left and right circular polarization components of the electric field respectively.

\subsubsection{Transition dipole moments}
One can define linear and circular transition dipole moment as in \cite{lourenco-martins_optical_2021}

\begin{equation}
    \left\{\begin{split}
        \hat{d}_x = -\frac{e w_0}{2} (\hat{a}_x^{} + \hat{a}_x^\dag) \\
        \hat{d}_y = -\frac{e w_0}{2} (\hat{a}_y^{} + \hat{a}_y^\dag)
    \end{split} \right.
\end{equation}

\begin{equation}
    \left\{\begin{split}
        \hat{d}_\circlearrowleft = -\frac{e w_0}{2} (\hat{a}_\circlearrowleft^{} + \hat{a}_\circlearrowright^\dag) \\
        \hat{d}_\circlearrowright = -\frac{e w_0}{2} (\hat{a}_\circlearrowright^{} + \hat{a}_\circlearrowleft^\dag)
    \end{split} \right.
\end{equation}

Note that the anihiliation of a left quantum of angular momentum induces the same transition as the creation of a right quantum of angular momentum. Conservation of angular momentum is here explicit : the creation of a left quantum of angular momentum in the electron transverse wavefunction induces a right circular field in the target, Resulting in zero total change of angular momentum.


\subsection{Quadrupolar term}
From \cite{novotny_principles_2011}, expressing the Quadrupolar term of the Hamiltonian

\begin{equation}
    \hat{V}_2 = - [\hat{\overleftrightarrow{Q}} \Vec{\nabla}] \cdot \mathbf{E}
\end{equation}
Where the quadrupolar tensor is expressed as
\begin{equation}
    \hat{\overleftrightarrow{Q}} = \frac{1}{2}  q\ \hat{\textbf{r}} \otimes \hat{\textbf{r}}
\end{equation}
With $q = -e$ for electrons
\begin{equation} \label{Q_tensor}
    \hat{Q}_{\nu \mu} = -\frac{e}{2} \hat{r}_\nu \hat{r}_\mu
\end{equation}
Which can be made in a 2D representation of the Quadrupolar tensor
\begin{equation}
    \hat{\overleftrightarrow{Q}} = -\frac{e}{2} \left[ \begin{matrix}
\hat{x}^2 & \hat{x}\hat{y} \\
\hat{y}\hat{x} & \hat{y}^2 
\end{matrix} \right]
\end{equation}
Considering only one electron we get
\begin{equation}\label{eq_si:V_Q_pos}
    \hat{V}_2 = \frac{1}{2} e \left[ \hat{x}^2 \partial_x E_x + \hat{y}^2 \partial_y E_y + \hat{x}\hat{y} ( \partial_x E_y + \partial_y E_x) \right]
\end{equation}
We remind that we consider implicit the $(\textbf{R}_0,q_z,\omega)$ dependency for the electric field operators.
\textbf{Nota bene : }The quadrupolar tensor can be found expressed under different ways,
\begin{equation} \label{Q_tensor_wrong}
    Q_{\nu \mu} = -\frac{e}{2} (r_\nu r_\mu - \textbf{r}^2 \delta^{}_{\nu \mu})
\end{equation}
Which is not equivalent to (\ref{Q_tensor}) since we are working in the xy-plane and $\partial_x E_x + \partial_y E_y  = -i q_z E_z$ in vacuum (Maxwell Gauss).

\subsubsection{Quadrupolar potential : Hermite-Gauss}
Injecting (\ref{x_expression_circlearrowleftadder_op_}) and (\ref{y_expression_circlearrowleftadder_op_}) in (\ref{eq_si:V_Q_pos}), we obtain
\begin{equation}
    \hat{V}_2 = \frac{e w_0^2}{8} \left[ \partial_x E_x (\hat{a}_x^{} + \hat{a}_x^\dag)^2 + \partial_y E_y (\hat{a}_y^{} + \hat{a}_y^\dag)^2 + (\partial_x E_y + \partial_y E_x) (\hat{a}_x^{} +\hat{a}_x^\dag) (\hat{a}_y^{} +\hat{a}_y^\dag) \right]
\end{equation}
Which can be developped into
\begin{equation}
\begin{split}
    \hat{V}_2 = \frac{e w_0^2}{8} \left[ \partial_x E_x (\hat{a}_x^2 + \hat{a}_x^{\dag 2} + \{\hat{a}_x,\hat{a}_x^\dag\}) + \partial_y E_y (\hat{a}_y^2 + \hat{a}_y^{\dag 2} + \{\hat{a}_y,\hat{a}_y^\dag\}) \right. \\ \left. + (\partial_x E_y + \partial_y E_x) (\hat{a}_x^{} +\hat{a}_x^\dag) (\hat{a}_y^{} +\hat{a}_y^\dag) \right]
\end{split}
\end{equation}
With $\{\hat{a}_x, \hat{a}_x^\dag \}$ being the anti-commutator. We note that $\{\hat{a}_x, \hat{a}_x^\dag \} = \mathds{1} + 2 \hat{N}_x$
\begin{equation}\label{Quadrupolar_Hermite_dev}
\begin{split}
    \hat{V}_2 = \frac{e w_0^2}{8} \left[ \partial_x E_x (\hat{a}_x^2 + \hat{a}_x^{\dag 2} + 2 \hat{N}_x + \mathds{1}) + \partial_y E_y (\hat{a}_y^2 + \hat{a}_y^{\dag 2} + 2 \hat{N}_y + \mathds{1}) \right. \\ \left. + (\partial_x E_y + \partial_y E_x) (\hat{a}_x^{} \hat{a}_y^{}  + \hat{a}_x^{} \hat{a}_y^\dag + \hat{a}_x^\dag \hat{a}_y^{}  + \hat{a}_x^\dag \hat{a}_y^\dag) \right]
\end{split}
\end{equation}

\subsubsection{Quadrupolar potential : Laguerre-Gauss}

Expressing the quadrupolar term with circular ladder operators by using (\ref{left a/c op}) and (\ref{right a/c op}): 

\begin{equation}\label{V_Q_LG_condensed_corrected}
    \hat{V}_2 = \frac{e w_0^2}{8} \left[ (\hat{a}_\circlearrowleft^{} + \hat{a}_\circlearrowright^\dag)^2 \partial_\circlearrowright E_\circlearrowleft + (\hat{a}_\circlearrowright^{} + \hat{a}_\circlearrowleft^\dag)^2 \partial_\circlearrowleft E_\circlearrowright +\frac{1}{2} \{ \hat{a}_\circlearrowright^{} + \hat{a}_\circlearrowleft^\dag , \hat{a}_\circlearrowleft^{} + \hat{a}_\circlearrowright^\dag \} (\partial_\circlearrowright E_\circlearrowright + \partial_\circlearrowleft E_\circlearrowleft) \right]
\end{equation}
Or expanded
\begin{equation}\label{Quadrupolar_Laguerre_dev}
    \begin{split}
    \hat{V}_2 =   \frac{e w_0^2}{8} \left[ (\hat{a}_\circlearrowleft^2 + \hat{a}_\circlearrowright^{\dag 2} + 2 \hat{a}_\circlearrowleft^{} \hat{a}_\circlearrowright^\dag) \partial_\circlearrowright E_\circlearrowleft + (\hat{a}_\circlearrowright^2 + \hat{a}_\circlearrowleft^{\dag 2} + 2 \hat{a}_\circlearrowright^{} \hat{a}_\circlearrowleft^\dag) \partial_\circlearrowleft E_\circlearrowright \right.\\ \left.+ (\hat{a}_\circlearrowright^{} \hat{a}_\circlearrowleft^{} + \hat{a}_\circlearrowright^\dag \hat{a}_\circlearrowleft^\dag + \mathds{1} + \hat{N}_\circlearrowright + \hat{N}_\circlearrowleft) (\partial_\circlearrowright E_\circlearrowright + \partial_\circlearrowleft E_\circlearrowleft) \right]
\end{split}
\end{equation}
Where we used the \textbf{Wirtinger derivatives} defined here as
\begin{equation}
    \begin{split}
        \partial_\circlearrowright = \frac{\partial}{\partial \circlearrowright} = \frac{1}{\sqrt{2}}\left( \frac{\partial}{\partial x} + i\frac{\partial}{\partial y}\right) \\ 
        \partial_\circlearrowleft = \frac{\partial}{\partial \circlearrowleft} = \frac{1}{\sqrt{2}}\left( \frac{\partial}{\partial x} - i\frac{\partial}{\partial y} \right)
    \end{split}
\end{equation}
These derivative acting on $E_\circlearrowright$ and $E_\circlearrowleft$ in their cartesian decomposition give
\begin{equation}
\begin{split}
    \partial_\circlearrowright E_\circlearrowleft = \frac{1}{2}(\partial_x +i\partial_y) (E_x +i E_y) = \frac{1}{2} (\partial_x E_x - \partial_y E_y +i \partial_x E_y + i \partial_y E_x) \\
    \partial_\circlearrowleft E_\circlearrowright = \frac{1}{2}(\partial_x -i\partial_y) (E_x -i E_y) = \frac{1}{2} (\partial_x E_x - \partial_y E_y -i \partial_x E_y - i \partial_y E_x) 
\end{split}
\end{equation}

\begin{equation}
    \partial_\circlearrowleft E_\circlearrowleft + \partial_\circlearrowright E_\circlearrowright = \frac{1}{2} (\partial_x - i\partial_y) (E_x +i E_y) +  \frac{1}{2} (\partial_x + i\partial_y) (E_x - i E_y) = \partial_x E_x + \partial_y E_y 
\end{equation}

\subsection{Higher orders}
We don't treat explicitly higher orders of the multipolar development in this text. They are quite easily accessible through application of the above formalism. They may reveal interesting properties to be probed. Considering the dipolar moment to be the first order of the development, then for any $k>1$
\begin{equation}
    \hat{V}_k = - \frac{q}{k!}\hat{\textbf{r}} \cdot [\hat{\textbf{r}} \cdot \nabla]^{k-1} \textbf{E}
\end{equation}
Development of higher orders under the ladder operator formalism get cumbersome but one could think of using a formal calculation algorithm.

\section{Shaped EELS interaction computation scheme}

We now have the expression for the PSEELS probability (\ref{EELS_multipolar_simplified}), the decomposition in two different basis of any transverse state for the electron beam through a combination of ladder operators (\ref{eq_si:HG_ladder_op},\ref{eq_si:LG_ladder_op}) and finally the expression of every order of the projected potential with ladder operators. We have all the tools to compute the transition probability between any initial and final electron transverse state. We here give as an example the computations for orders (0),(1),(2).

\subsection{Transition amplitudes and probability}
To simplify the notations, we will consider the amplitude $g_{0,i\rightarrow j}$ associated to a transition between order i and j beams. The corresponding EELS probabilities are the squared modulus of these amplitudes. To the leading order, one gets:

\begin{equation} \label{Amplitude_HG_general_si}
    g_{0,i\rightarrow j} = \sum_{k} \mathcal{M}^{(k)}[\hat{V}_k]\;\delta_{k,\vert i-j \vert}
\end{equation}
where $\mathcal{M}^{(k)}$ corresponds to the transition of order $k$ involving only the multipolar term $\hat{V}_k$. $i$ and $j$ are the initial and final transverse electron profile orders as defined in section (3.5). $\delta_{k,\vert i-j \vert}$ is the Kronecker delta, defined as 
\begin{equation}\label{Kronecker_delta}
    \delta_{a,b} = \left\{\begin{matrix}
 1& \text{if }\ a=b  \\
 0& \text{if }\ a\neq b
\end{matrix}\right.
\end{equation}
There are non zero contributions to $g_{0,i\rightarrow j}$ from higher order terms in the potential $k>|i-j|$. We will show in section (5.2) that they are negligible under this approximation scheme and that (\ref{Amplitude_HG_general_si}) an excellent approximation.\\

We add a superscript to the $\mathcal{M}$ amplitudes to specify which basis is considered (HG or LG) and a subscript to indicate the indices of the initial and final transverse profiles. We choose to remove the $[\hat{V}_k]$ dependence as it is redundant with the $(k)$ superscript indicating the order of the $\mathcal{M}$ term. As an example, $\mathcal{M}_{20,10}^{(1)HG}$ is the amplitude of the transition $\vert HG_{20} \rangle \rightarrow \vert HG_{10} \rangle$ given by the dipolar term of the potential. The general amplitudes read:
\begin{equation}
    \mathcal{M}_{n_x n_y, n_x' n_y'}^{(k)HG} = \frac{1}{\hbar v} \langle HG_{n_x', n_y'} \vert \hat{V}_k \vert HG_{n_x, n_y} \rangle
\end{equation}

\begin{equation}
    \mathcal{M}_{n_\circlearrowright n_\circlearrowleft, n_\circlearrowright' n_\circlearrowleft'}^{(k)LG} = \frac{1}{\hbar v} \langle LG_{n_\circlearrowright', n_\circlearrowleft'} \vert \hat{V}_k \vert LG_{n_\circlearrowright, n_\circlearrowleft} \rangle
\end{equation}

\subsubsection{Computation example}
To ease the comprehension, we compute step by step an example: Considering the transition $\vert HG_{20} \rangle \rightarrow \vert HG_{10} \rangle$. It is a order $|i-j|=|(2+0)-(1+0)|=1$ process. The transition amplitude $g_{0,2\rightarrow 1}$ is given to leading order by $\mathcal{M}_{20,10}^{(1)HG}$. We compute the amplitude:

\begin{equation}
    \mathcal{M}_{20, 10}^{(1)HG} = \frac{1}{\hbar v} \langle HG_{1, 0} \vert \hat{V}_1 \vert HG_{2,0} \rangle
\end{equation}
The procedure to compute explicitly this transition amplitude is now direct by using (\ref{Dipolar_Hermite}), applying the ladder operators on the states (\ref{HG_ladder_rules}) to reach the point where no more ladder operators are present and only remain brackets of HG or $LG$ states, being orthogonal sets they are then trivial to compute.
\begin{equation}
    \mathcal{M}_{20, 10}^{(1)HG} = \frac{e w_0}{2 \hbar v} \langle HG_{1, 0} \vert \big[(\hat{a}^{}_x+\hat{a}^\dag_x)E_x +(\hat{a}^{}_y+\hat{a}^\dag_y)E_y\big] \vert HG_{2,0} \rangle
\end{equation}
And 
\begin{equation}
    \langle HG_{1, 0} \vert \hat{a}^{}_y \vert HG_{2,0} \rangle = \langle HG_{1, 1}  \vert HG_{2,0} \rangle = 0 = \langle HG_{1, 0} \vert \hat{a}^{\dag}_y \vert HG_{2,0} \rangle
\end{equation}
\begin{equation}
    \langle HG_{1, 0} \vert ( \hat{a}^{}_x+\hat{a}^\dag_x ) \vert HG_{2,0} \rangle = \sqrt{2}\underset{=\ 1}{\underbrace{\langle HG_{1, 0} \vert HG_{1,0} \rangle}} + \sqrt{3}\underset{=\ 0}{\underbrace{\langle HG_{1, 0} \vert HG_{3,0} \rangle}} = \sqrt{2}
\end{equation}
Leading to 
\begin{equation}
     \mathcal{M}_{20, 10}^{(1)HG} = \frac{e w_0}{\sqrt{2} \hbar v} E_{x} 
\end{equation}
And so 
\begin{equation}
\boxed{
\Gamma^{HG}_{20\rightarrow10} (\textbf{R}_0,\omega, q_z) = \Big| \frac{e w_0}{\sqrt{2} \hbar v} E_{n0,x}(\textbf{R}_0,\omega, q_z)  \Big|^2
}
\end{equation}
Where we brought back all the implicit dependencies. The intensity of the transition from a $\vert HG_{20} \rangle$ to a $\vert HG_{10} \rangle$ is proportional to the squared absolute value of the x component of the transition electric field. In fact any transition probability will be proportional to the squared absolute value of a linear combination of electric field components or derivatives. At order $k=1$, the electron behaves as a small dipole measuring the electric near-field at the nano scale. This shows that the electron beam effectively behaves as a nanoscale source of polarized white light, enabling a prolongation of the optics experiment results to the deep sub-wavelength scale. This property named Optical Polarization Analog is well discussed in recent literature \cite{lourenco-martins_optical_2021,bourgeois_optical_2023}.\\

Following this analytical scheme, the contributions to the different orders of $g_0$ can be computed for any initial and final states in both basis. In the following we list the general results. The selection rules are made explicit by the Kronecker delta found in the resulting amplitudes. The interested reader wanting to use the computation formalism is advised to refer to the present section rather than the general expressions in a first approach.
\subsubsection{Electrostatic term}

\begin{equation}
    \hat{V}_0  = \hat{\rho}. \Phi  = -e \hat{n} \Phi
\end{equation}
With $\hat{n}$ the electron number operator, equivalent to identity when considering a single electron as we do. We compute the transition amplitude between any two Hermite-Gauss transverse states
\begin{equation}
    \mathcal{M}_{n_x n_y, n_x' n_y'}^{(0)HG} = \frac{-e}{\hbar v} \langle HG_{n_x',n_y'} \vert HG_{n_x,n_y} \rangle \Phi 
\end{equation}
Resulting in 
\begin{equation}
    \mathcal{M}_{n_x n_y, n_x' n_y'}^{(0)HG} = \frac{-e}{\hbar v} \delta^{}_{n_x,n_x'} \delta^{}_{n_y,n_y'} \Phi 
\end{equation}
With $\delta^{}_{n_x,n_x'}$ the Kronecker delta. And for Laguerre-Gauss transverse states
\begin{equation}
    \mathcal{M}_{n_\circlearrowright n_\circlearrowleft, n_\circlearrowright' n_\circlearrowleft'}^{(0)LG} = \frac{-e}{\hbar v} \delta^{}_{n_\circlearrowright,n_\circlearrowright'} \delta^{}_{n_\circlearrowleft,n_\circlearrowleft'} \Phi
\end{equation}
As found before, the $0^{th}$ order of PSEELS is equivalent to classical EELS experiments. It corresponds to no change in the transverse wavefunction $\Delta n_x =\Delta n_y =\Delta n_\circlearrowleft =\Delta n_\circlearrowright = 0$.

\subsubsection{Dipolar Hermite transitions}
Studying now order $k=1$ transitions carried by the dipolar term of the projected potential. 
\begin{equation}
    \hat{V}_1 = \frac{e w_0}{2} \left[ (\hat{a}_x^{} + \hat{a}_x^\dag) E_x  +  (\hat{a}_y^{} + \hat{a}_y^\dag) E_y  \right]
\end{equation}
The general amplitude reads:
\begin{equation}
\begin{split}
    \mathcal{M}_{n_x n_y, n_x' n_y'}^{(1)HG} = \frac{e w_0}{2 \hbar v} [(\sqrt{n_x}\delta^{}_{n_{x-1},n_x'}\delta^{}_{n_y,n_y'} + \sqrt{n_x +1}\delta^{}_{n_{x+1},n_x'}\delta^{}_{n_y,n_y'}) E_x  \\+ (\sqrt{n_y}\delta^{}_{n_{x},n_x'}\delta^{}_{n_{y-1},n_y'} + \sqrt{n_y +1}\delta^{}_{n_{x},n_x'}\delta^{}_{n_{y+1},n_y'}) E_y ]
\end{split}
\end{equation}
We get that the only non zero amplitudes, hence the only transitions carrying a linear dipolar moment are those where the transverse wavefunction index changes by one unit exactly : $\Delta n_x = \pm 1$ or $\Delta n_y = \pm 1$. The measured field has the direction of the ladder operator implied in the transition, this is linear momentum conservation.

\subsubsection{Dipolar Laguerre transitions}
For Laguerre-Gauss type transverse wavefunctions, the computations are similar.
\begin{equation}\label{Dipolar_Laguerre_si}
    \hat{V}_1 = \frac{e w_0}{2} \left[ (\hat{a}_\circlearrowleft^{} + \hat{a}_\circlearrowright^\dag) E_\circlearrowleft   +  (\hat{a}_\circlearrowright^{} + \hat{a}_\circlearrowleft^\dag) E_\circlearrowright    \right]
\end{equation}

\begin{equation}
    \begin{split}
        \mathcal{M}_{n_\circlearrowright n_\circlearrowleft, n_\circlearrowright' n_\circlearrowleft'}^{(1)LG} = \frac{e w_0}{2 \hbar v} \left[E_\circlearrowright \Big(\sqrt{n_\circlearrowright}\delta^{}_{n_\circlearrowright-1,n_\circlearrowright'}\delta^{}_{n_\circlearrowleft,n_\circlearrowleft'} + \sqrt{n_\circlearrowleft +1}\delta^{}_{n_\circlearrowright,n_\circlearrowright'}\delta^{}_{n_\circlearrowleft+1,n_\circlearrowleft'}\Big)  \right. \\ \left.+ E_\circlearrowleft \Big(\sqrt{n_\circlearrowright +1}\delta^{}_{n_\circlearrowright+1,n_\circlearrowright'}\delta^{}_{n_\circlearrowleft,n_\circlearrowleft'} + \sqrt{n_\circlearrowleft}\delta^{}_{n_\circlearrowright,n_\circlearrowright'}\delta^{}_{n_\circlearrowleft-1,n_\circlearrowleft'} \Big) \right]
    \end{split}
\end{equation}
Again, the non zero amplitudes are those for which $\Delta n_\circlearrowleft = \pm 1$ or $\Delta n_\circlearrowright = \pm 1$. Interestingly, for the interaction amplitude to be proportional to $E_\circlearrowleft$, the left circular component of the electric field, the transverse profile of the electron needs to either lose one quantum of left circular momentum or gain one quantum of right circular momentum : it is the conservation of angular momentum.
\subsubsection{Quadrupolar Hermite transitions}
We get to order $k=2$ transition carried by the quadrupolar potential expressed in Hermite-Gauss basis:
\begin{equation}
\begin{split}
    \hat{V}_2 = \frac{e w_0^2}{8} \left[ \partial_x E_x (\hat{a}_x^2 + \hat{a}_x^{\dag 2} + 2 \hat{N}_x + \mathds{1}) + \partial_y E_y (\hat{a}_y^2 + \hat{a}_y^{\dag 2} + 2 \hat{N}_y + \mathds{1}) \right. \\ \left. + (\partial_x E_y + \partial_y E_x) (\hat{a}_x^{} \hat{a}_y^{}  + \hat{a}_x^{} \hat{a}_y^\dag + \hat{a}_x^\dag \hat{a}_y^{}  + \hat{a}_x^\dag \hat{a}_y^\dag) \right]
\end{split}
\end{equation}
The optical nearfield quantity probed does not change when probing $\vert HG_{n_x n_y}\rangle \rightarrow \vert HG_{n_x' n_y'}\rangle$ or $\vert HG_{n_x' n_y'}\rangle \rightarrow \vert HG_{n_x n_y}\rangle$. The general amplitude reads:

\begin{equation}
    \begin{split}
        \mathcal{M}_{n_x n_y, n_x' n_y'}^{(2)HG} = \frac{e w_0^2}{8 \hbar v} \left[ \partial_x E_x \delta^{}_{n_y,n_y'} \Big(\sqrt{n_x (n_x-1)}\delta^{}_{n_x-2,n_x'} + \sqrt{(n_x+1)(n_x+2)}\delta^{}_{n_x+2,n_x'} + (2n_x +1)\delta^{}_{n_x,n_x'}\Big) \right. \\ 
        \left. \partial_y E_y \delta^{}_{n_x,n_x'} \Big(\sqrt{n_y (n_y-1)}\delta^{}_{n_y-2,n_y'} + \sqrt{(n_y+1)(n_y+2)}\delta^{}_{n_y+2,n_y'} + (2n_y +1)\delta^{}_{n_y,n_y'}\Big) \right. \\
        \left. + (\partial_x E_y + \partial_y E_x)\Big(\sqrt{n_x n_y} \delta^{}_{n_x-1,n_x'}\delta^{}_{n_y-1,n_y'} + \sqrt{n_x (n_y +1)} \delta^{}_{n_x-1,n_x'}\delta^{}_{n_y+1,n_y'} \right.\\ \left.+ \sqrt{(n_x + 1) n_y} \delta^{}_{n_x+1,n_x'}\delta^{}_{n_y-1,n_y'} + \sqrt{(n_x +1) (n_y +1)} \delta^{}_{n_x+1,n_x'}\delta^{}_{n_y+1,n_y'}\Big)\right]
    \end{split}
\end{equation}
This term of interaction can give direct access to nearfield quantities such as $\partial_x E_x$ or $\partial_x E_y + \partial_y E_x$ corresponding to quadrupolar signature in the nearfield. Due to the mixed terms $a_x a_y^\dagger$, we can measure quadrupole systems while keeping first order shaping at the price of complex post-selection ($\vert HG_{10} \rangle \rightarrow \vert HG_{01} \rangle$).

\subsubsection{Quadrupolar Laguerre transitions}
Also order $k=2$ transition but relying on quadrupolar potential expressed in Laguerre-Gauss basis
\begin{equation}\label{Quadrupolar_Laguerre_dev_2}
    \begin{split}
    \hat{V}_2 =   \frac{e w_0^2}{8} \left[ (\hat{a}_\circlearrowleft^2 + \hat{a}_\circlearrowright^{\dag 2} + 2 \hat{a}_\circlearrowleft^{} \hat{a}_\circlearrowright^\dag) \partial_\circlearrowright E_\circlearrowleft + (\hat{a}_\circlearrowright^2 + \hat{a}_\circlearrowleft^{\dag 2} + 2 \hat{a}_\circlearrowright^{} \hat{a}_\circlearrowleft^\dag) \partial_\circlearrowleft E_\circlearrowright \right.\\ \left.+ (\hat{a}_\circlearrowright^{} \hat{a}_\circlearrowleft^{} + \hat{a}_\circlearrowright^\dag \hat{a}_\circlearrowleft^\dag + \mathds{1} + \hat{N}_\circlearrowright + \hat{N}_\circlearrowleft) (\partial_\circlearrowright E_\circlearrowright + \partial_\circlearrowleft E_\circlearrowleft) \right]
\end{split}
\end{equation}
We get to complex expressions when keeping them general
\begin{equation}
    \begin{split}
        \mathcal{M}_{n_\circlearrowright n_\circlearrowleft, n_\circlearrowright' n_\circlearrowleft'}^{(2)LG} =  \frac{e w_0^2}{8 \hbar v} &\Big[ \ \ \partial_\circlearrowleft E_\circlearrowright \Big(  \sqrt{n_\circlearrowright (n_\circlearrowright -1)} \delta^{}_{n_\circlearrowright-2, n_\circlearrowright'}\delta^{}_{n_\circlearrowleft, n_\circlearrowleft'} + \sqrt{(n_\circlearrowleft+1 )(n_\circlearrowleft+2)}\delta^{}_{n_\circlearrowright, n_\circlearrowright'}\delta^{}_{n_\circlearrowleft+2, n_\circlearrowleft'} \\
         & \hspace{10ex} + 2 \sqrt{n_\circlearrowright(n_\circlearrowleft+1)}\delta^{}_{n_\circlearrowright-1, n_\circlearrowright'}\delta^{}_{n_\circlearrowleft+1, n_\circlearrowleft'} \Big) \\
         &+\partial_\circlearrowright E_\circlearrowleft \Big(\sqrt{n_\circlearrowleft (n_\circlearrowleft -1)} \delta^{}_{n_\circlearrowright, n_\circlearrowright'}\delta^{}_{n_\circlearrowleft-2, n_\circlearrowleft'} + \sqrt{(n_\circlearrowright+1 )(n_\circlearrowright+2)}\delta^{}_{n_\circlearrowright+2, n_\circlearrowright'}\delta^{}_{n_\circlearrowleft, n_\circlearrowleft'}\\ 
         & \hspace{10ex }+ 2 \sqrt{(n_\circlearrowright+1)n_\circlearrowleft}\delta^{}_{n_\circlearrowright+1, n_\circlearrowright'}\delta^{}_{n_\circlearrowleft-1, n_\circlearrowleft'}\Big) \\
         &  + (\partial_\circlearrowright E_\circlearrowright + \partial_\circlearrowleft E_\circlearrowleft) \Big( \sqrt{n_\circlearrowright n_\circlearrowleft} \delta^{}_{n_\circlearrowright-1, n_\circlearrowright'}\delta^{}_{n_\circlearrowleft-1, n_\circlearrowleft'} + \sqrt{(n_\circlearrowright+1) (n_\circlearrowleft+1)} \delta^{}_{n_\circlearrowright+1, n_\circlearrowright'}\delta^{}_{n_\circlearrowleft+1, n_\circlearrowleft'} \\
         & \hspace{21ex }+ (1 + n_\circlearrowright + n_\circlearrowleft)\delta^{}_{n_\circlearrowright, n_\circlearrowright'}\delta^{}_{n_\circlearrowleft, n_\circlearrowleft'}\Big) \Big]
    \end{split}
\end{equation}
The Quadrupolar term of the potential is used to access three quantities : $\partial_\circlearrowleft E_\circlearrowright$;$\partial_\circlearrowright E_\circlearrowleft$;$\partial_\circlearrowright E_\circlearrowright + \partial_\circlearrowleft E_\circlearrowleft$.

\subsubsection{Transition amplitudes to Gaussian final state}
As an example, we specify the previous general transition amplitudes for the case of a gaussian final state $\vert G \rangle$. The transition amplitudes can be reformulated as
\begin{equation}
    \mathcal{M}_{n_x n_y, 00}^{(1)HG} = \frac{e w_0}{2 \hbar v} \Big[ \delta_{n_x,1} \delta_{n_y,0} E_x + \delta_{n_x,0} \delta_{n_y,1} E_y \Big]
\end{equation}
\begin{equation}
    \mathcal{M}_{n_\circlearrowleft n_\circlearrowright, 00}^{(1)LG} = \frac{e w_0}{2 \hbar v} \Big[ \delta_{n_\circlearrowleft,1} \delta_{n_\circlearrowright,0} E_\circlearrowleft + \delta_{n_\circlearrowleft,0} \delta_{n_\circlearrowright,1} E_\circlearrowright \Big]
\end{equation}
\begin{equation}
    \mathcal{M}_{n_x n_y, 00}^{(2)HG} = \frac{e w_0^2}{8 \hbar v} \Big[ \sqrt{2}\partial_x E_x \delta_{n_x,2} \delta_{n_y,0} + \sqrt{2}\partial_y E_y \delta_{n_x,0} \delta_{n_y,2} + (\partial_x E_y + \partial_y E_x) \delta_{n_x,1} \delta_{n_y,1} + (\partial_x E_x + \partial_y E_y) \delta_{n_y,0} \delta_{n_y,0} \Big]
\end{equation}
\begin{equation}
    \mathcal{M}_{n_\circlearrowright n_\circlearrowleft, 0 0}^{(2)LG} =  \frac{e w_0^2}{8 \hbar v} \Big[ \sqrt{2}\partial_\circlearrowleft E_\circlearrowright \delta_{n_\circlearrowleft,0} \delta_{n_\circlearrowright,2} + \sqrt{2}\partial_\circlearrowright E_\circlearrowleft \delta_{n_\circlearrowleft,2} \delta_{n_\circlearrowright,0} + (\partial_\circlearrowleft E_\circlearrowleft + \partial_\circlearrowright E_\circlearrowright)(\delta_{n_\circlearrowleft,1}\delta_{n_\circlearrowright,1} + \delta_{n_\circlearrowleft,0}\delta_{n_\circlearrowright,0}) \Big]
\end{equation}

\subsection{Higher $k$ orders of potential impact on lower $|i-j|$ order transitions amplitudes}
Some transitions have contributions from multiple orders of the potential, once the correct $|i-j|=k$ number of creation/anihiliation operator leads to a non-zero transition amplitude, adding a $\hat{a}\hat{a}^\dag$ gives a non zero contribution but corresponds to a higher order $k+2$ of potential implied in the transition. Each order comes with a multiplication by $w_0$ and a derivative of the electric field operator. Since the typical variation length of the electric field is much larger than $w_0$, this contribution negligible in the transition amplitude. For instance the $\vert G \rangle \rightarrow \vert G \rangle$ gets a $\mathcal{M}_{00, 00}^{(2)HG} = \frac{e w_0^2}{8 \hbar v} (\partial_x E_x + \partial_y E_y)$ contribution from the quadrupolar term of the potential. One can check that in the narrow beam limit 
\begin{equation}
    \Big|\mathcal{M}_{00, 00}^{(2)HG}\Big| \ll \Big|\mathcal{M}_{00, 00}^{(0)HG}\Big|
\end{equation}
This shows that (\ref{Amplitude_HG_general_si}) is an excellent approximation and can be considered exact.

\subsection{Multi mode EELS}
Reminding the PSEELS rate (\ref{eq_si:EELS_qproba_g0}), we have
\begin{equation}
\begin{split}
    \Gamma^\text{PSEELS}(\omega) & = \tilde{\sum_n}  \left| \frac{1}{\hbar v}  \langle \Psi_{f,\perp} \vert \langle n \vert \hat{V}_\text{proj} \vert 0 \rangle \vert \Psi_{i,\perp} \rangle \right|^2 \\ & = \tilde{\sum_n} | g_{0,n} |^2 
\end{split}
\end{equation}
The measured PSEELS interaction probability contains contributions from the different photonic states $n$ that can be populated through the interaction. This is an 'incoherent' sum, the states are probed independently from each other, the total PSEELS interaction can be computed mode by mode, making our single mode computations before perfectly correct.

\subsection{Identifying physical quantities measured under these techniques}
Proper pre and post selection of the electron transverse wavefunction can be used to choose precisely which quantity of the Near-field to probe. For instance : 

\begin{equation}
    g_{0,11\rightarrow00}^{HG} \propto  \partial_x E_y + \partial_y E_x
\end{equation}
\begin{equation}
    g_{0,20\rightarrow00}^{HG} \propto  \partial_x E_x
\end{equation}
We draw tables of the probed quantity for each transition for HG beams and for LG beams, presented in the main text.

\section{Mapping}
Retrieving the dependencies, the transition amplitude $g_{0,i\rightarrow j} (\textbf{R}_0,q_z,\omega)$ is proportional to nearfield quantities at a given position $\textbf{R}_0$, position of the electron beam. Scanning the electron beam over the sample will give a map of the considered quantity.

\subsection{Another point of view : maps and spectra}
A map is measured at a precise energy. We can also fix the position of the electron beam and measure a spectrum for a choosen transition. The resulting spectrum gives the symetry-matching modes excited in the sample as done experimentally in \cite{guzzinati_probing_2017}. The PSEELS setup can be seen both as a tool to map certain nearfield quantities or as a tool to precisely know the symetry of the populated mode in the sample, e.g exciting on demand quadrupolar transitions in the sample.

\bibliography{References.bib} 
\bibliographystyle{unsrt}

\end{document}